\documentclass{article}

\usepackage[preprint, nonatbib]{neurips_2026}

\usepackage{marvosym}
\usepackage{fontawesome}
\usepackage{enumitem}
\usepackage{subcaption}
\usepackage[utf8]{inputenc} 
\usepackage[T1]{fontenc}    
\usepackage{hyperref}       
\usepackage{url}            
\usepackage{booktabs}       
\usepackage{amsfonts}       
\usepackage{nicefrac}       
\usepackage{microtype}      
\usepackage{ltablex}
\keepXColumns

\usepackage{amssymb}     
\usepackage{xspace}
\usepackage{graphicx}
\usepackage{booktabs}
\usepackage{multirow}
\usepackage{listings,upquote,soul,framed}
\usepackage[table,xcdraw, dvipsnames]{xcolor}
\usepackage{caption}
\usepackage{array}
\usepackage{tabularx}
\usepackage{longtable}
\usepackage{adjustbox}
\usepackage{markdown}
\markdownSetup{
  pipeTables = true
}
\usepackage{pifont}
\usepackage[normalem]{ulem}
\usepackage{wrapfig}
\usepackage{amssymb}

\usepackage[most]{tcolorbox}
\usepackage{fvextra}
\usepackage{pifont}
\usepackage{float}
\usepackage{seqsplit}

\DefineVerbatimEnvironment{PromptVerbatim}{Verbatim}{%
   fontsize=\scriptsize,
   breaklines=true,
   breakanywhere=true,
   breaksymbolleft={},
   breaksymbolright={},
   tabsize=2
}
\newcommand{\repo}[1]{\texttt{\seqsplit{#1}}}

\newcommand{\llm}{\textsc{LLM}\xspace}

\newcommand{\bfit}[1]{\textbf{\textit{#1}}}
\newcommand{\cmark}{\ding{51}}
\newcommand{\xmark}{\ding{55}}

\newcommand{\oursys}{\textsc{SetupX}\xspace}
\newcommand{\ourproblem}{functionality-correct repository setup\xspace}
\newcommand{\Ourproblem}{Functionality-correct repository setup\xspace}

\newcommand{\xpu}{\bfit{XPU}\xspace}
\newcommand{\xpudb}{\bfit{XPU Database}\xspace}

\newcommand{\benchmark}{our benchmark\xspace}

\title{\oursys: Can LLM Agents Learn from Past Failures in Functionality-Correct Code Repository Setup?}

\author{
    Zihang Zhou$^{1}$, 
    Ziqian Ren$^{1}$,  
    Yukai Wu$^{2}$, 
    Yingjie Xiong$^{1}$,
    \textbf{Wei Zhou$^{1}$,}\\
    \textbf{Chao Peng$^{3}$,}
    \textbf{Dong Zhang$^{4}$,}
    \textbf{Bingheng Yan$^{4}$,}
    \textbf{Xuanhe Zhou \Letter $^{1}$,} 
    \textbf{Fan Wu$^{1}$}\\
    $^1$ Shanghai Jiao Tong University 
    $^2$ Beijing University of Posts and Telecommunications\\
    $^3$ Independent Researcher 
    $^4$ Jinan Inspur Data Technology Co., Ltd.\\
    \texttt{zzh2024@sjtu.edu.cn}
}

\begin{document}


\maketitle



\begin{abstract}
\Ourproblem~aims to configure execution environments (e.g., dependencies, build scripts) to successfully execute a repository's documented features. 
It presents significant challenges due to diverse, repository-specific failures, including dependency incompatibilities, missing toolchains, incomplete installations, and verification-strategy mismatches.
Existing LLM agents struggle to robustly resolve these issues, specifically failing to support (1) cross-repository experience transfer, (2) multi-step trial-and-repair under non-invertible state changes, and (3) robust verification of setup outcomes to distinguish setup-induced failures from repository bugs.
To address this, we introduce \oursys, an experiential learning-based setup framework.
First, we construct a \emph{Self-Evolving Experience Representation} (\xpu), a dual-modality knowledge unit encoding setup signals, textual guidance, executable actions to dynamically transfer verified environment fixes to unseen repositories.
Second, we employ \emph{Experience-Augmented Speculative Execution} backed by a LIFO Docker snapshot stack, enabling the agent to proactively trial fixes and safely roll back to known-good states. 
Third, we introduce a \emph{Prosecutor-Judge Verification Protocol} that separates evidence collection from final judgment, enabling more reliable setup verification beyond superficial build-time metrics.
Evaluation results on carefully-crafted benchmarks show \oursys achieves highest performance (e.g., 92\% pass rate) and outperforms the strongest baseline by over 19\%. Crucially, \oursys excels in complex multi-repository setup requiring coordinating multiple interconnected services across different containers.  The code repository is available at\  \textcolor{blue}{{\url{https://github.com/OpenDataBox/SetupX}}}.
\end{abstract}

\vspace{-1.25em}
\section{Introduction}
\label{sec:introduction}
\vspace{-.5em}

Large language model (LLM)-based agents demonstrate impressive performance on complex software engineering tasks, such as feature implementation and issue resolution~\cite{metagpt,sweagent}.
However, most benchmarks over these tasks assume pre-configured environments~\cite{jimenez2024swebench, deveval}, while real-world deployment requires building executable environments from scratch (i.e., functionality-correct repository setup).
This process is notoriously complex, involving the resolution of intricate dependency conflicts, the installation of missing build tools, and the reconciliation of version incompatibilities.

\begin{figure}[t]
    \centering
    \includegraphics[width=0.98\linewidth]{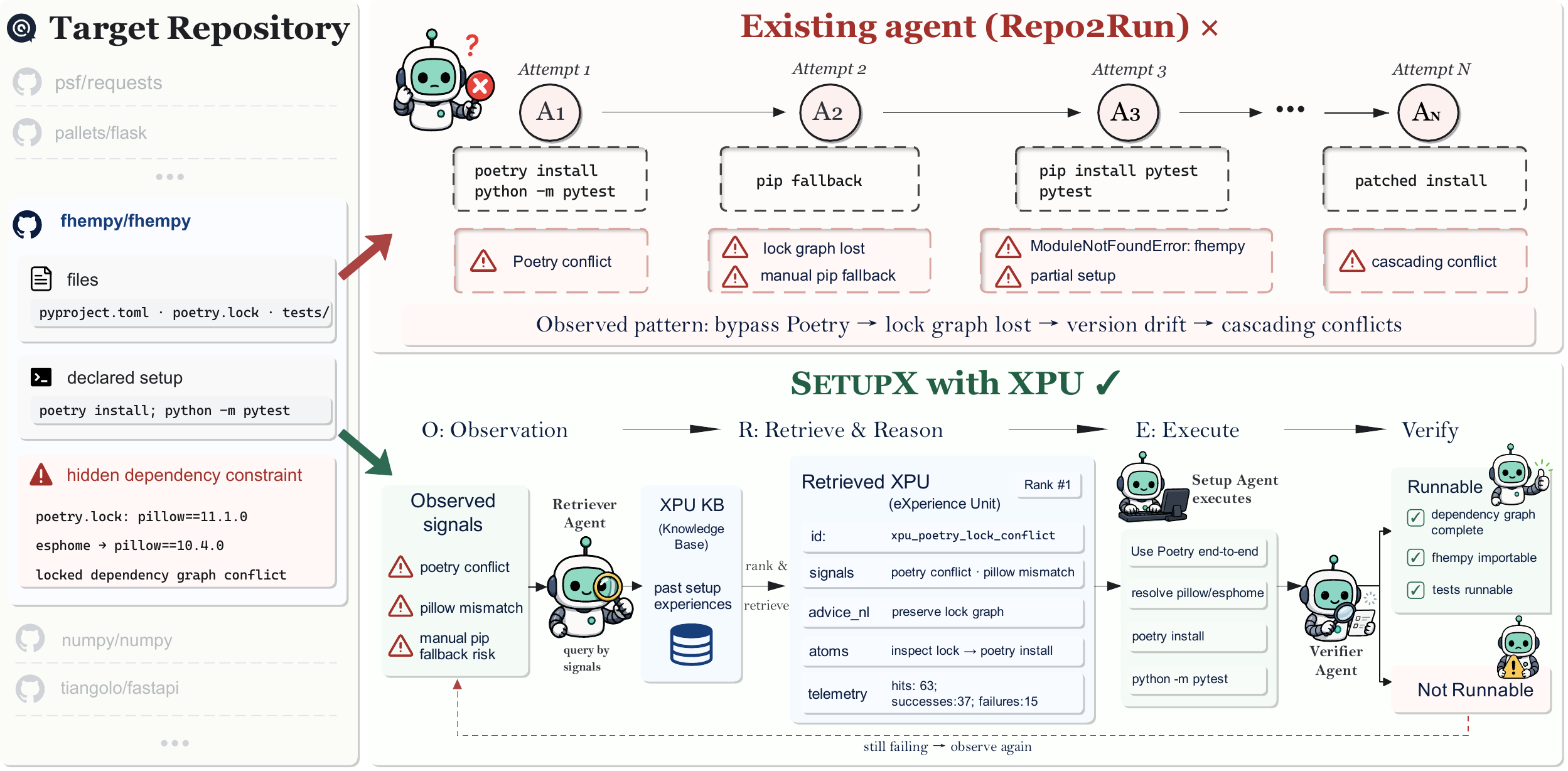}
    \caption{Comparison between existing setup agents (e.g.,  Repo2Run~\cite{hu2026reporun}) and \oursys.}
    \label{fig:introduction}
    \vspace{-.75em}
\end{figure}

Existing LLM-based coding agents (e.g., Claude Code~\cite{claudecode}, OpenHands~\cite{wang2024openhands}) and specialized setup tools (e.g., Repo2Run~\cite{hu2026reporun}, ExecutionAgent~\cite{executionagent}) show promise in automating this process. 
However, they can make significant setup mistakes in many scenarios. For instance, as shown in Figure~\ref{fig:introduction}, when configuring \textsc{fhempy}~(\textit{Python extension for smart-home automation}), Repo2Run~\cite{hu2026reporun} enters a stateless retry loop: after \texttt{poetry install} fails due to a locked dependency conflict, the agent repeats similar commands and falls back to \texttt{pip install}, causing version drift and cascading dependency errors without extracting reusable diagnostic evidence.


Moving beyond the \textsc{fhempy} failure example, we identify three fundamental challenges for effective automatic setup:
\ding{182} First, \emph{Cross-Repository Failure Patterns:} Similar dependency conflicts recur across repositories, while agents like ExecutionAgent~\cite{executionagent} treat each repository as an isolated episode, discarding transferable setup experience. 
\ding{183} Second, \emph{Non-Invertible State Changes:} A failed installation can incur package conflicts that simple undo operations cannot reverse, corrupting the environment and preventing naive trial-and-error from succeeding. 
\ding{184} Finally, \emph{Unreliable Success Signals:} There is no unified, trustworthy criterion for setup completion. Current systems either keep setup, execution and success assessment within the same agent-driven workflow (e.g., Claude Code~\cite{claudecode}, ExecutionAgent~\cite{executionagent}), or rely on shallow objective proxies (e.g., Repo2Run's~\cite{hu2026reporun} \texttt{pytest}-oriented test-execution feedback and EnvBench's~\cite{envbench} static missing-import checks) that cannot reflect functional usability or attribute failures to their root cause. Setup quality thus remains fundamentally unverifiable from within the agent itself.
Collectively, these challenges motivate our research question: \emph{Can a unified agentic framework systematically transfer cross-repository experience, safely explore complex setup states, and reliably verify setup outcomes within a single continuous loop?}


To answer this question, we introduce \oursys, an experience-driven framework for autonomous repository setup. In summary, we make the following four contributions:

\noindent $\bullet$ \textbf{(1) Self-Evolving Experience Representation.}
To exploit cross-repository failure patterns, we design the \emph{eXPerience Unit} (\xpu), a structured knowledge representation that jointly encodes diagnostic signals, natural-language advice, and executable atomic operations. Supported by a two-layer retrieval mechanism and an anchor-based delayed audit, the \xpu database continuously self-curates based on real deployment outcomes, ensuring robustness to evolving knowledge.

\noindent $\bullet$ \textbf{(2) Experience-Augmented Speculative Execution.}
\oursys retrieves fixes from the \xpu database and trials them under \emph{speculative execution}, where container snapshots enable rollback to any prior known-good state, supporting safe multi-step exploration of non-invertible repairs.

\noindent $\bullet$ \textbf{(3) Prosecutor-Judge Verification Protocol.} 
To overcome unreliable success signals, we propose a two-phase verification mechanism that structurally separates execution from evaluation. An independent Prosecutor issues evidence-backed charges against the configured environment, while a Judge independently verifies each charge before rendering a final verdict. This structural separation reliably distinguishes setup-induced failures from inherent repository defects.

\noindent $\bullet$ \textbf{(4) Empirical Validation.}
We evaluate \oursys on a curated 100-repository Python benchmark drawn from the EnvBench challenge pool~\cite{envbench}. Under unified independent adjudication, \oursys achieves a 92\% pass rate, outperforming the strongest general-purpose agent (Claude Code) by 19\% and the strongest specialized tool (ExecutionAgent) by 33\%. A 22-family multi-repository evaluation further confirms \oursys's capability to generalize across complex deployment scenarios.

\vspace{-.5em}
\section{Problem Definition}
\label{sec:problem}
\vspace{-.5em}

Prior works~\cite{hu2026reporun,executionagent} primarily verify the configured environment by executing the project's test suite, leaving documented user-facing commands unchecked. In this work, we consider a stricter setting, termed \emph{functionality-correct repository setup}.
Given a code repository $\mathcal{R}=(\mathcal{P}, S)$, where $\mathcal{P}$ denotes the source code and $S$ specifies a commit, let $\mathcal{F}_{\mathcal{R}}$ denote the set of functionality-oriented execution targets declared by the repository (e.g., test suite, documented commands). The goal is to construct a valid execution environment $\mathcal{E}$ that (1) is realized as a reproducible container image produced either by building from a Dockerfile (as adopted by prior methods such as Repo2Run~\cite{hu2026reporun}) or by committing an interactively configured container, and (2) executing the targets in $\mathcal{F}_{\mathcal{R}}$ on this image does not produce any setup-induced failure:
\begin{equation}
    \mathrm{Valid}(\mathcal{E}, \mathcal{R}) \iff
    \mathcal{E} \in \mathcal{D} \;\land\;
    \forall f \in \mathcal{F}_{\mathcal{R}},\
    \mathrm{Run}(f,\mathcal{E}) \notin \mathcal{C}_{\mathrm{fail}}
\end{equation}
where $\mathcal{D}$ denotes the space of reproducible container images, and $\mathcal{C}_{\mathrm{fail}}$ denotes the set of setup-induced failures, which we empirically group into four recurring categories: \emph{dependency incompatibility}, \emph{toolchain missing}, \emph{invalid or incomplete package installation}, and \emph{verification-strategy mismatch}.
Note that failures caused by bugs in the repository's source code, flawed test logic, or unsupported functionality are treated as project-intrinsic failures and are not considered in this work.

To address the three challenges for effective setup, \oursys (Figure~\ref{fig:overview}) relies on a self-curating knowledge base of \emph{eXPerience Units} (\xpu), which is continuously expanded by distilling offline execution trajectories for cross-repository deployments (Section~\ref{sec:xpu_system}).
To safely apply this knowledge within the irreversible container space $\mathcal{D}$, the agent trials retrieved \xpu via snapshot-backed \emph{speculative execution} (Section~\ref{sec:setup_agent}).
Finally, an adversarial Prosecutor-Judge protocol rigorously audits the configured environment to detect any remaining setup-induced failures $\mathcal{C}_{\mathrm{fail}}$ (Section~\ref{sec:verification}).

\section{Self-Evolving Experience Representation}
\label{sec:xpu_system}

\begin{figure*}[!t]
    \centering
    \includegraphics[width=.9\textwidth]{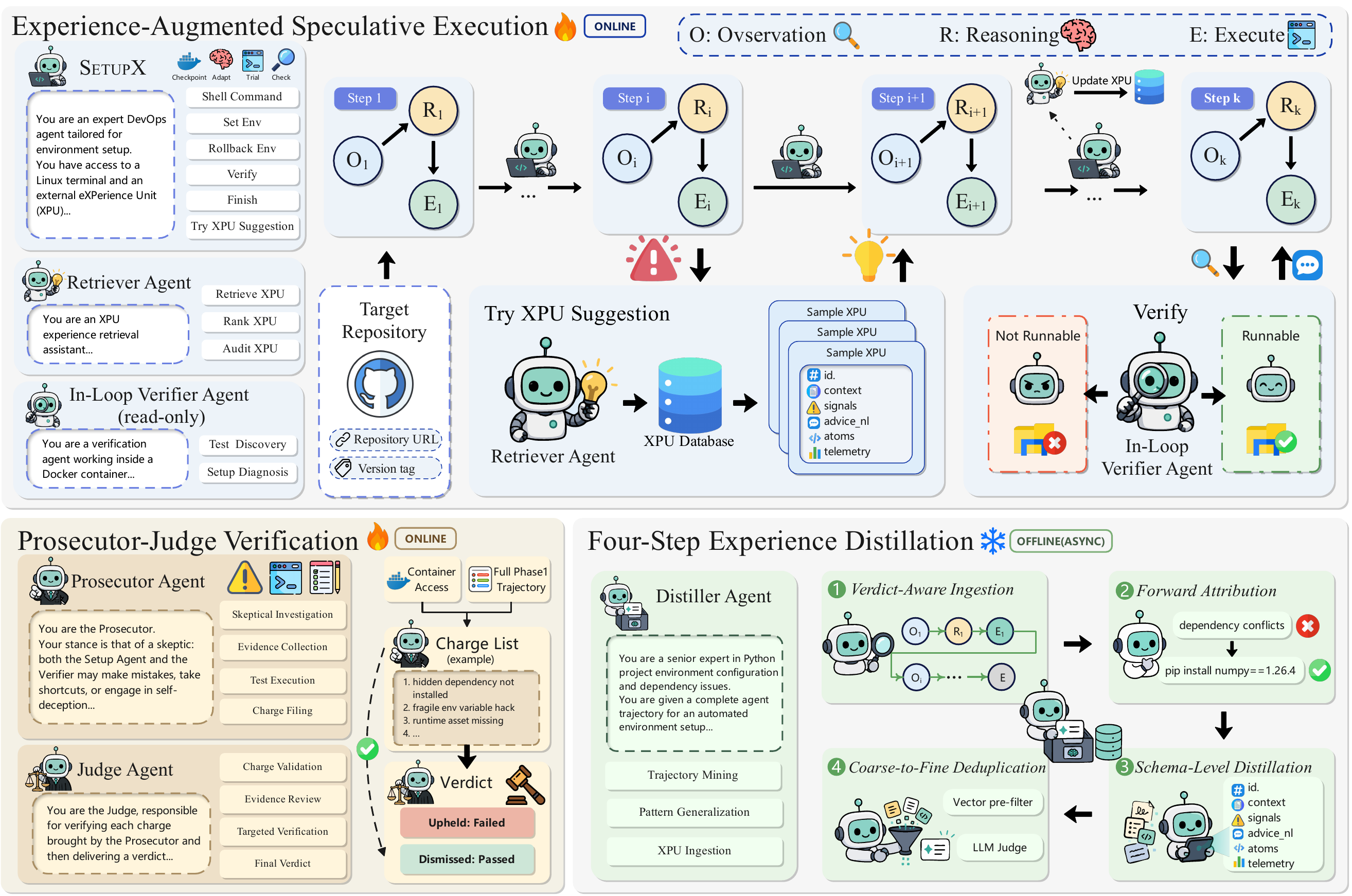}
    \caption{Overview of \oursys. We perform the environment setup in a ReAct loop with \xpu-guided speculative execution and in-loop verification, where the Prosecutor-Judge verification independently audits the result  transferable lessons are ingested into the \xpudb.}
    \vspace{-1em}
    \label{fig:overview}
\end{figure*}

Existing memory-augmented agents struggle to efficiently accumulate and transfer setup experience, often resorting to storing verbose, non-generalizable trajectories. 
Moreover, {natural-language distillations}~\cite{reflexion,zhao2024expel} store text advice that is hard to search and track for historical success, and {executable skill libraries}~\cite{wang2023voyager,chen2026skillcraft} provide runnable code but lack structured indexes for error matching.
To resolve this, \oursys~maintains cross-repository setup experience to a self-evolving database: (1) the \emph{Structured eXPerience Unit (\xpu)}, a formalized representation mapping setup failures to illustrative advice and executable operations;
(2) an offline \emph{Four-Step Experience Distillation} pipeline that expands the \xpu base by transforming raw trajectories into deduplicated records;
and (3) an online \emph{Anchored-Based Delayed Audit} that calibrates each \xpu entry based on real deployment feedback.

\noindent \underline{$\blacktriangleright$ \textbf{(1) Structured eXPerience Unit (\xpu).}}
To effectively bridge reasoning with execution, provide exact error context, and track historical success, we propose the \emph{eXPerience Unit} (\xpu), a structured knowledge record that natively integrating diagnostic context, natural language guidance, and executable code.
Formally, an \xpu is defined as: $\text{\xpu} = \langle \textit{id},\;\textit{signals},\; \textit{advice\_nl},\; \textit{atoms},\; \textit{telemetry} \rangle$,
where \textit{signals} indexes the exact error context for precise retrieval, including applicability conditions (e.g., OS, build tools) and specific error messages;
\textit{advice\_nl} and \textit{atoms} form a dual representation to link reasoning with action, explaining the root cause in natural language and encoding the fix as executable operations, respectively;
and \textit{telemetry} enables online self-curation by tracking a deployment record ($\langle \textit{hits}, \textit{successes}, \textit{failures} \rangle$) that the delayed audit updates and the retriever consumes.



\noindent \underline{$\blacktriangleright$ \textbf{(2) Four-Step Experience Distillation.}}
To systematically incorporate new \xpu entries and drive continuous evolution, we propose an offline \llm-based distillation process that evaluates every execution trajectory and its past evaluations through a four-step pipeline (Figure~\ref{fig:overview}):
\emph{\ding{182} Verdict-Aware Ingestion:} A Distiller Agent analyzes the trajectory using evaluation signals, extracting corrective insights from actions, rather than relying exclusively on the agent's self-reported successes;
\emph{\ding{183} Forward Attribution:} Since error recovery is rarely linear and the true impact of a fix frequently emerges steps later, an attribution pass localizes concrete environment problems (e.g. dependency conflicts, build-tool failures) and traces them forward to the specific actions that resolved them;
\emph{\ding{184} Schema-Level Distillation:} An \llm maps each problem-fix pair into the \xpu schema, abstracting the transferable reasoning pattern instead of merely recording exact commands;
\emph{\ding{185} Coarse-to-Fine Deduplication:} A vector pre-filter (cosine similarity $\geq 0.85$) identifies potential duplicates for an \llm judge to verify semantic equivalence; confirmed duplicates trigger a merge (unioning signal sets and fusing advice), while all remaining candidates are ingested as novel entries.

\begin{figure}[!t]
    \centering
    \includegraphics[width=0.85\columnwidth]{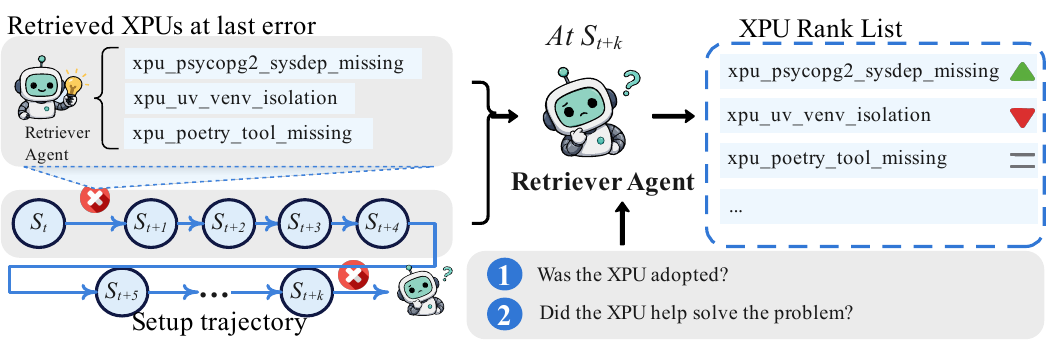}
    \caption{Delayed audit feedback loop. When a new retrieval is triggered, the Retriever Agent first audits the previous \xpu recommendations against the subsequent setup trajectory, assigns verdicts, and updates per-\xpu telemetry that feeds back into future retrieval ranking.}
    \label{fig:delayed_audit}
    \vspace{-.25em}
\end{figure}

\noindent \underline{$\blacktriangleright$ \textbf{(3) Anchor-Based Delayed Audit.}}
To further reliably assess \xpu quality and capture delayed feedback (where real impact surfaces steps later when downstream dependencies resolve), we implement an online anchor-based retrospective evaluation.
As shown in Figure~\ref{fig:delayed_audit}, the mechanism operates by recording the recommended \xpu IDs and the current execution history position at every retrieval step.
When the next retrieval occurs, the Retriever Agent audits the intervening execution steps to classify each prior recommendation as a \textit{success}, \textit{failure}, or \textit{neutral}.
These classifications continuously update the telemetry of each \xpu, which dictates future retrieval rankings via tier assignment. The complete evaluation prompt and update rules are available in Appendix~\ref{app:delayed-audit}.

\section{Experience-Augmented Speculative Execution}
\label{sec:setup_agent}

Once \oursys has access to reusable \xpu{}s, the next challenge is to apply them safely inside
a mutable container. To address this, \oursys augments the standard ReAct paradigm~\cite{yao2023react} with three mechanisms: (1)~\emph{Agentic Experience Retrieval}
(Section~\ref{sec:retriever}); (2)~\emph{Speculative Execution} (Section~\ref{sec:speculative}); (3)~\emph{In-Loop Verifier Agent} (Section~\ref{sec:inloop_verifier}).
At each step, \oursys first observes the container and the last action's output;
it automatically invokes the Retriever Agent to identify relevant \xpu{}s based on setup failures, which are added to the prompt. 
\oursys then reasons over recent history together with the retrieved natural-language guidance
and executable atoms, and selects one of six actions (provided in Appendix~\ref{sec:action_space}), where three are specialized for experience-driven setup: \texttt{TRY\_XPU\_SUGGESTION}, \texttt{ROLLBACK\_ENV}, and \texttt{VERIFY}. The setup loop terminates when \oursys emits the \texttt{FINISH} action.

\subsection{Agentic Experience Retrieval}
\label{sec:retriever}

Naive embedding-based search over raw error messages yields low retrieval precision since syntactically similar errors may have entirely different root causes (e.g., \texttt{ModuleNotFoundError} for a missing system library \emph{vs.}\ a missing pip package), while semantically related problems may produce syntactically different error text.
To address this, \oursys uses a dedicated \emph{Retriever Agent} that operates in an independent context, shielding the main agent's prompt from retrieval noise. The Retriever Agent constructs a \emph{hybrid situation} (an LLM-generated
current state summary paired with the raw command output and error text) and runs
coarse-to-fine retrieval against the knowledge base.


\textbf{Coarse-to-Fine Retrieval.}
We first perform vector coarse filtering: the hybrid situation is embedded via \texttt{text-embedding-3-small} and queried against a pgvector index.
The top-$N$ ($N{=}10$) candidates are ranked by a composite score:
\begin{equation}
    s_{\text{composite}} = s_{\text{sim}} \cdot (1 + r_{\text{success}}) \cdot b_{\text{tier}},
    \qquad
    b_{\text{tier}} =
    \begin{cases}
        1.5, & \text{if } \textit{hits} \!\geq\! 5 \text{ and } r_{\text{success}} \!\geq\! a \quad (\textit{Golden}), \\
        0.6, & \text{if } \textit{hits} \!\geq\! 5 \text{ and } r_{\text{success}} \!<\! b \quad (\textit{Cold}), \\
        1.0, & \text{otherwise} \quad (\textit{Normal}).
    \end{cases}
    \label{eq:composite}
\end{equation}
where $s_{\text{sim}}$ is cosine similarity, $r_{\text{success}} = \textit{successes} / \max(\textit{hits}, 1)$ is the historical success rate, and $b_{\text{tier}}$ is a tier boost. We set the thresholds to $a{=}0.2$ and $b{=}0.1$ based on current \xpu telemetry, yielding a roughly balanced split across the three tiers while assigning entries with insufficient evidence to \textit{Normal}. This score combines semantic relevance with empirical utility, so experiences with high similarity but low success rates are naturally deprioritized. Finally, an LLM refines the ranking by reading the full content and selecting the top-$K$ most relevant entries, where $K{=}3$.


\vspace{-.5em}
\subsection{Speculative Execution}
\label{sec:speculative}
\vspace{-.5em}

Environment repair is \emph{non-monotonic}: an earlier fix may appear successful but
introduce a latent conflict that surfaces only several steps later, leaving the
offending step buried beneath subsequent attempts. To safely explore candidate fixes
under such delayed feedback, we give the agent \emph{active}, multi-step control
over the trajectory through a LIFO snapshot stack $\mathcal{S}$ exposed via two
actions.

\ding{182} \textbf{\texttt{TRY\_XPU\_SUGGESTION}} pushes a checkpoint of the current container
onto $\mathcal{S}$ via Docker's copy-on-write \texttt{docker commit}, then trials
the adapted \xpu commands:
\begin{equation}
    \mathcal{C}_{t+1} =
    \begin{cases}
        \mathcal{C}'_t & \text{if } \texttt{trial}(\mathcal{C}'_t) = \textit{success}, \\
        \texttt{restore}(\mathcal{S}.\texttt{pop}()) & \text{otherwise}.
    \end{cases}
    \label{eq:speculative}
\end{equation}
where the new state is retained over success trial, otherwise the just-pushed snapshot is restored
in place. The full five-stage trial protocol is provided in Appendix~\ref{sec:try_xpu}.


\ding{183} \textbf{\texttt{ROLLBACK\_ENV}} lets the agent \emph{actively} pop any number of
frames off $\mathcal{S}$ and return the container to \emph{any} earlier known-good
checkpoint, not only the latest one. Once subsequent diagnostics reveal that several
past attempts have collectively steered the trajectory into a dead end, the agent
invokes \texttt{ROLLBACK\_ENV} to retreat past all of them and explore an
alternative path.

Unlike Repo2Run's command-level adaptive rollback~\cite{hu2026reporun} which restores the environment to the pre-execution state of a failed command, \oursys exposes rollback as an agent-controlled LIFO snapshot stack, allowing entire sub-paths to be abandoned once latent conflicts surface.

\subsection{In-Loop Verifier Agent}
\label{sec:inloop_verifier}
The In-Loop Verifier Agent is a lightweight read-only ReAct agent invoked when \oursys issues \texttt{VERIFY}.
It operates within the same container under strict constraints:
\emph{(1)~Read Only}, prohibited from installing packages, modifying environment variables, or altering project files;
\emph{(2)~Structured Protocol}, fixed sequence of structural reconnaissance, test suite location, test execution via the project's native runner, and failure analysis;
\emph{(3)~Dual-Outcome Semantics}, distinguishes setup-induced failures (actionable) from inherent project limitations (non-blocking for \texttt{FINISH}).
The result is a per-step, attribution-aware diagnostic signal that lets \oursys drive targeted repair rather than blind retry.

\section{Prosecutor-Judge Verification Protocol}
\label{sec:verification}

Existing verification falls into two modes, neither providing both independent adjudication and per-failure attribution: \emph{self-reported verification} ~\cite{claudecode,wang2024openhands} lets the agent adjudicate its own success, conflating execution with adjudication; \emph{non-attributable external verification}~\cite{envbench,hu2026reporun} collapses the pipeline into a binary verdict that cannot separate setup defects from project intrinsics.
To address this, we propose a \emph{Prosecutor-Judge protocol}, an independent post-hoc audit structurally decoupled from the setup that pairs an investigative
\emph{Prosecutor} with an adjudicative \emph{Judge} to expose premature
success masked by surface signals.

\textbf{Prosecutor Agent.}
The Prosecutor operates as a skeptical investigator with full container access, following a forced multi-step investigation protocol designed to prevent superficial assessments:
(1)~identify the project's language and build system by examining marker files;
(2)~review the full trajectory during setup;
(3)~verify that core dependencies are importable using the same interpreter \oursys configured, providing a test-independent verification signal that remains meaningful even for repositories with minimal or absent test suites;
(4)~consult the README for documented entry points (e.g., example scripts, CLI commands) and verify a representative subset launches without setup-induced failure;
(5)~run the test suite independently and compare results;
(6)~formulate specific charges backed by concrete evidence, following the principle ``when in doubt, prosecute''.

\textbf{Judge Agent.} The Judge receives the Prosecutor's charges and performs \emph{targeted verification}: for each charge, it executes 1--2 independent commands to confirm or refute the evidence.
A charge is dismissed if the evidence is contradicted, the issue concerns optional dependencies, or the failure is caused by external factors.
If any charge is upheld, the setup is ruled \emph{guilty}; and \emph{not\_guilty} otherwise.



This separation creates an effective check: the Prosecutor finds possible failures, while the Judge verifies them independently, making the final verdict more reliable.

\vspace{-.5em}
\section{Experiments}
\label{sec:experiments}
\vspace{-.5em}

\subsection{Benchmark Construction}
\label{sec:benchmark}

Our evaluation is built on the EnvBench Python challenge pool, which contains 329 Python repositories~\cite{envbench}. We focus on Python because its dependency ecosystem is large and complex, its packaging practices are heterogeneous, and it frequently involves native-library interactions, all of which make automated setup particularly challenging~\cite{pypi_empirical,watchman}.
From this repository pool, we construct \benchmark{}, a curated benchmark of 100 repositories for systematic comparison. The construction proceeds in three steps: we first retain repositories with detectable test assets; then run a preliminary agent-based trial to stratify repositories by setup difficulty; finally, sample across application domains and project maturity to obtain the final benchmark.
For example, \texttt{lark} is a low-difficulty project because it has minimal dependencies and a straightforward installation process. In contrast, \texttt{acme} is a high-difficulty project because it involves optional ecosystems such as TensorFlow and JAX, with strict version coupling that causes the agent to fail even after multiple setup retries.

\begin{figure}[!t]
  \centering
  \includegraphics[width=\linewidth]{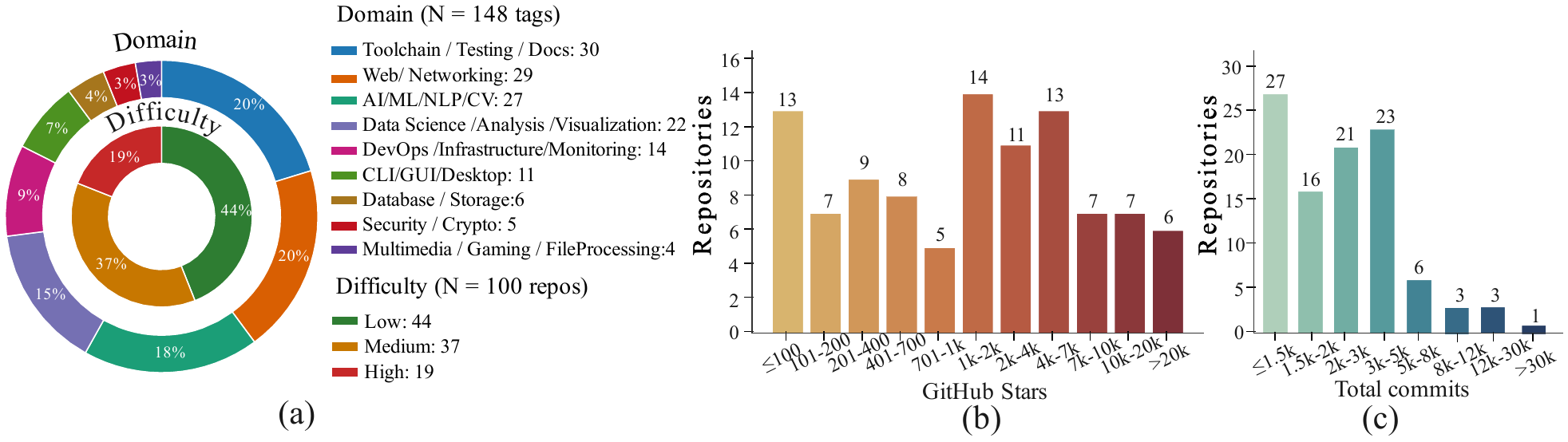}
  \caption{Statistics of The Setup-Specific Benchmark.}
  \label{fig:benchmark_composition}
  \vspace{-.5em}
\end{figure}

The final benchmark contains 44 low-, 37 medium-, and 19 high-difficulty repositories. It covers ten major application areas and 148 domain tags, and each repository is pinned to a fixed revision. As shown in Figure~\ref{fig:benchmark_composition}, the benchmark covers diverse application domains and project scales.

We also construct a multi-repository track with 22 composite scenarios, covering four representative patterns: client--server, plugin--host, multi-service deployment, and platform--SDK. Because these tasks require workflow-level evidence across repositories, we validate them through manual review and use this track as a complementary stress test. The full rubric is placed in Appendix \ref{app:rubric}. Each family is rated on a four-tier scale (Full / Mostly / Partial / Shallow) based on a five-check rubric covering clone integrity, editable install, module availability, pytest collection, and dependency consistency.

\subsection{Experimental Setup}
\label{sec:exp-setup}

 We evaluate eight systems on \benchmark{}: \oursys{} in two setup variants, with and without \xpu{}; three open-source CLI coding agents, Claude Code, Qwen Code, and OpenCode; and three
   specialized setup tools, ExecAgent, Repo2Run, and EnvBench. All systems use \texttt{qwen3.5-plus} as the backbone LLM for both setup execution and the unified \oursys{}
  prosecutor--judge adjudication, run inside a fresh \texttt{ubuntu:22.04} container with a global timeout of 3600 seconds for each run. The \xpu{} knowledge base used by \oursys{} is
  collected from the full EnvBench Python repository pool with 1536-dim \texttt{text-embedding-3-small} embeddings, and all entries corresponding to the 100 repositories in \benchmark{}
  are removed before evaluation. A run passes if the prosecutor files no accusation or the judge rejects all accusations; timeout or any \emph{guilty} verdict is counted as a failure.

\subsection{Main Results}
\label{sec:main_results}

\begin{figure}
    \centering
    \includegraphics[width=1\linewidth]{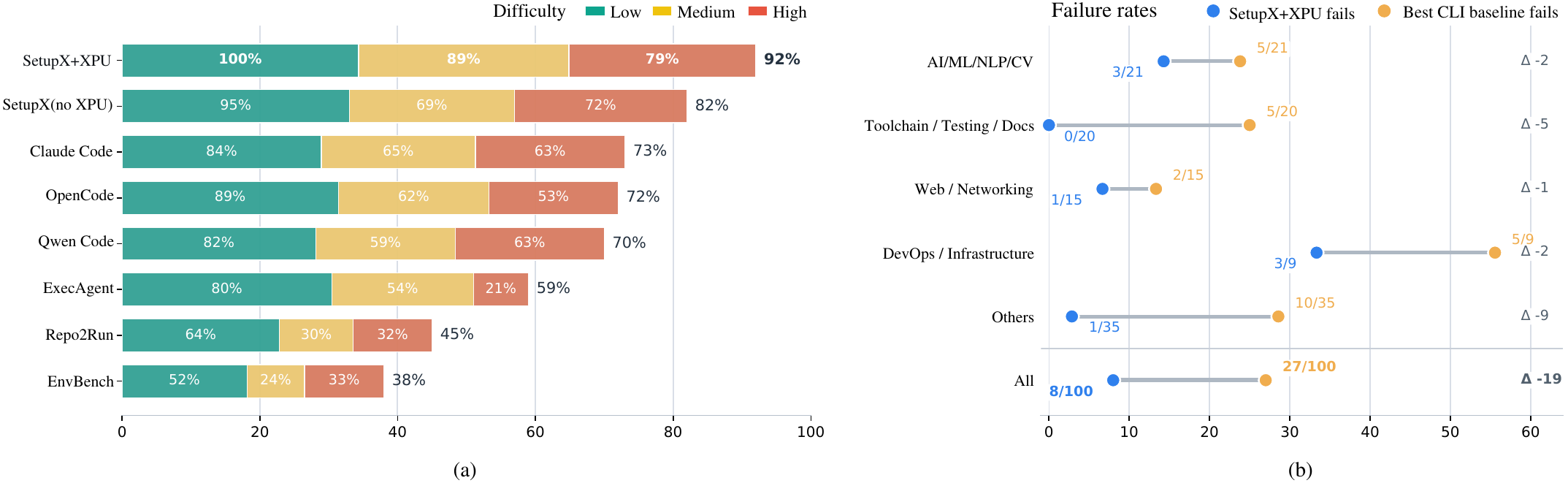}
    \caption{Main results on the single-repository track. (a) Setup pass rates by repository difficulty; (b) Domain-wise failure rates of SETUPX+XPU versus the best CLI baseline.}
    \label{fig:combined_figure}
    \vspace{-.75em}
\end{figure}

We first compare the setup pass rate of eight systems on \benchmark{} under the prosecutor--judge adjudication protocol. In parallel, we use the multi-repository family evaluation as a complementary stress test to examine whether setup methods can scale to deployment scenarios involving a host repository, component repositories, and sibling services. Figure~\ref{fig:combined_figure} reports pass rate on the single-repository benchmark, and Figure~\ref{fig:combined_figure} breaks down the failures by repository domain.

\oursys{}+\xpu{} achieves a 92\% pass rate on \benchmark{}, ranking first overall. It leads the strongest LLM agent, Claude Code, by 19 pp; the strongest specialized setup tool, ExecAgent, by 33 pp; and the remaining specialized setup tools by 47--54 pp. Stratified by \texttt{install\_complexity}, the advantage concentrates on medium- and high-difficulty repositories: on the high stratum, \oursys{}+\xpu{} reaches 79\%, outperforming Claude Code by 16 pp and \oursys{} (no \xpu{}) by 7 pp, while the low stratum is already close to the 100\% ceiling.

After removing \xpu{}, \oursys{} still reaches an 82\% pass rate. This decomposes the overall advantage of \oursys{} into two sources: the thought--action--verifier agent loop raises the pass rate to 82\%, and \xpu{} contributes an additional 10 pp on top of it. Thus, the advantage of \oursys{} is not entirely attributable to knowledge-base retrieval. We defer fine-grained attribution of \xpu{}'s internal components to Section~\ref{sec:ablation}.

Beyond the single-repository benchmark, we further compare \oursys{} with Qwen Code, a representative open-source CLI baseline, on 22 non-atomic families. Both agents reach Full or Mostly on 17 of the 22 families, but \oursys{} obtains substantially more Full verdicts (6 vs. 1). The gap concentrates at the Full--Mostly boundary: both agents can recognize and execute the cross-repository protocol layer (starting sibling services, deploying Kubernetes components, registering plugin entry points), but a Full verdict further requires each participating repository to be properly installed, importable, and consistent with its declared dependencies. Qwen Code's Mostly families repeatedly stall on setup-completeness issues—for example, the default \texttt{PATH} not being linked to the venv, missing \texttt{pytest}, or setup stopping at ``tests can be collected'' without actually running them. \oursys{} is more robust along this dimension; per-family verdicts and a representative case are provided in Appendix \ref{app:per-family-verdicts} and \ref{app:multi-repo_cases}.

Beyond overall pass rate and multi-repo verdicts, we inspect failure distributions on the 100-repository benchmark from two complementary perspectives. As shown in Figure~\ref{fig:combined_figure}, \oursys{}+\xpu{} has no more failures than the strongest CLI baseline in any domain, with the total failure count reduced from 27 to 8. The largest gap appears in Toolchain ($n=20$), where \oursys{}+\xpu{} has zero failures while the baseline has five—failures in this domain mainly involve missing build tools, compilers, and native libraries, where the toolchain fix patterns accumulated in the \xpu{} knowledge base provide the most direct benefit. The remaining \oursys{}+\xpu{} failures occur mainly in the AI/ML and DevOps domains, suggesting that retrieval can still be incomplete for repository-specific dependency stacks.

We further attribute failures by their underlying mechanisms. Table~\ref{tab:failure_taxonomy} reports the percentage of each system's failures involving C1--C4 categories. This taxonomy is derived from baseline failure-case analysis, and a single failure may involve multiple categories. C3 is the dominant failure type across all systems, at around 38\%, followed by C4 at around 25\%; together, the four categories cover the main failure modes of all systems.

\begin{table}[H]
\vspace{-.75em}
\centering
\small
\caption{Failure distribution across categories C1--C4 on the 100-repository benchmark. The C1--C4 taxonomy is derived from baseline failure analysis; failures spanning multiple categories are counted in each applicable row, so columns may sum to more than 100\%.}
\label{tab:failure_taxonomy}
\resizebox{\linewidth}{!}{%
\begin{tabular}{llrrr}
\toprule
Category & Description & Claude Code & Repo2Run & \oursys{}+\xpu{} \\
\midrule
C1 & Dependency / runtime version incompatibility & 22\% & 2\% & 38\% \\
C2 & Native / build toolchain gaps & 19\% & 7\% & 13\% \\
C3 & Invalid or incomplete package installation & 37\% & 40\% & 63\% \\
C4 & Verification strategy mismatch & 26\% & 25\% & 25\% \\
\bottomrule
\end{tabular}}
\end{table}

\subsection{Ablation Study}
\label{sec:ablation}

\begin{figure}[htbp]
  \centering
  \begin{subfigure}[b]{0.42\linewidth}
    \vspace{0pt}
    \centering
    \includegraphics[width=\linewidth]{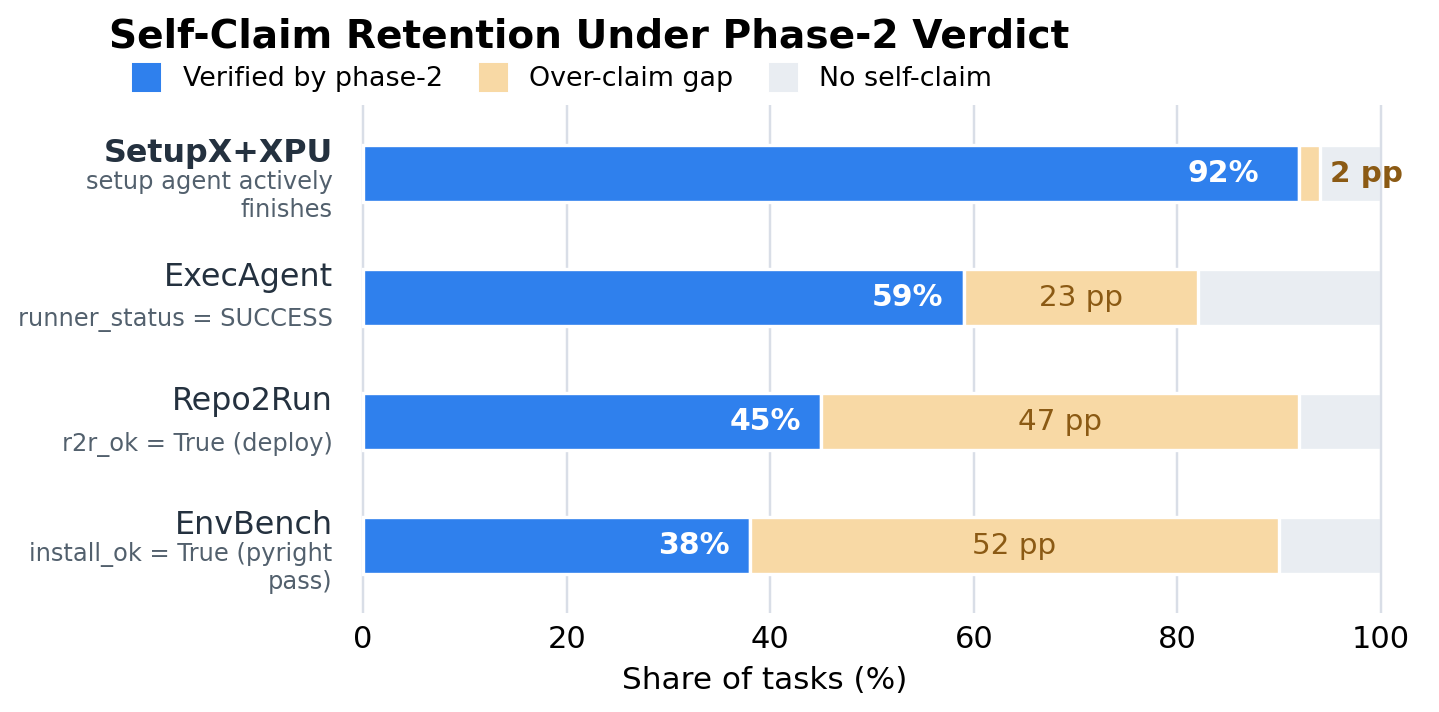}
    \caption{Self-claim vs. prosecutor--judge verdict.}
  \end{subfigure}
  \hfill
  \begin{subfigure}[b]{0.55\linewidth}
    \vspace{0pt}
    \centering
    \scriptsize
    \setlength{\tabcolsep}{2.2pt}
    \renewcommand{\arraystretch}{1.03}
    \resizebox{\linewidth}{!}{%
    \begin{tabular}{lccccrr}
      \toprule
      Setting & XPU & Rerank & KB & Backbone & Pass & Cost \\
      \midrule
      No-XPU & \xmark & -- & -- & qwen3.5-plus & 82 & 1.0$\times$ \\
      Selector + Clean $\star$ & \cmark & Selector & Clean & qwen3.5-plus & \textbf{92} & 1.0$\times$ \\
      Direct + Clean & \cmark & Direct & Clean & qwen3.5-plus & 86 & 1.0$\times$ \\
      Selector + Noisy & \cmark & Selector & Noisy & qwen3.5-plus & 88 & 1.0$\times$ \\
      Direct + Noisy & \cmark & Direct & Noisy & qwen3.5-plus & 81 & 1.0$\times$ \\
      Selector + Clean + Opus & \cmark & Selector & Clean & Claude Opus 4.6 & 92 & $\sim$10$\times$ \\
      \bottomrule
    \end{tabular}
    }
    \caption{Ablation settings.}
  \end{subfigure}
  \caption{Trustworthiness and ablation results.}
  \label{fig:trust_and_ablation}
\end{figure}

All pass-rate numbers in Section~\ref{sec:main_results} are based on the \oursys{} prosecutor--judge adjudication protocol. Before analyzing the \xpu{} retrieval pipeline, we first validate why this verdict standard is necessary. Figure~\ref{fig:trust_and_ablation}(a) juxtaposes the self-claim signals of three specialized setup tools, dockerfile build success, install command exit 0, and runner status code, with the unified prosecutor--judge verdict. The results show a systematic 23--52 pp gap between traditional tools' self-claims and our verdict. The root cause is that self-claim only checks that the pipeline did not crash: the dockerfile built, the install command returned 0, and the runner did not exit abnormally. None of these implies that the final setup is usable. EnvBench is the most representative example: 90\% of repositories pass \texttt{pyright} static analysis, but only 38\% pass when the prosecutor verifies dependency importability inside the container.

After fixing the verdict standard, we ablate the XPU retrieval pipeline along three dimensions. First, whether \xpu{} is enabled; second, whether retrieval is followed by LLM reranking, where Selector chooses 3 entries from the top-10 vector candidates and Direct returns the top-3 vector results directly; third, whether synthetic noise is injected into the \xpu{} knowledge base. In addition, we test the effect of scaling the backbone LLM by replacing \texttt{qwen3.5-plus} with Claude Opus 4.6 under the main Selector + Clean setting. Unless otherwise specified, all ablation settings use \texttt{qwen3.5-plus}, the 100-repository benchmark, and the prosecutor--judge adjudication protocol. See details in Appendix \ref{app:agent-prompts}. 
The ablation centers on the main setting Selector + Clean (\oursys{}+\xpu{}, 92\%). LLM reranking dominates the contributions: it brings a 6 pp gain on the Clean KB and a 7 pp gain on the Noisy KB. Substituting the Clean KB with its Noisy counterpart degrades all matched pairs, with Direct degrading the most because pure vector recall struggles to separate real advice from neighboring noise—reranking thus serves as a critical buffer for retrieval usability under noise. We further test whether a stronger backbone closes the residual gap: replacing \texttt{qwen3.5-plus} with Claude Opus 4.6 leaves pass rate unchanged at 92\%, while increasing LLM API cost by more than 10$\times$.\footnote{Based on per-million-token pricing as of May 2026: Claude Opus 4.6 at \$5.00 / \$25.00 (input / output) and \texttt{qwen3.5-plus} at \$0.40 / \$1.20.} Pipeline design and retrieval quality, rather than raw backbone capability, dominate at this scale, and the default setting offers a better cost--effectiveness tradeoff.

\section{Related Work}
\label{sec:related}


\noindent \underline{$\blacktriangleright$ \textbf{LLM Agents for Software Engineering.}}
\textbf{(1) Code-level repair agents.}
SWE-bench~\cite{jimenez2024swebench} established a dominant paradigm for repository-level program repair, motivating agentic systems such as SWE-agent~\cite{sweagent}, AutoCodeRover~\cite{autocoderover}, and OpenHands~\cite{wang2024openhands}.
These systems differ in interface design, localization strategy, autonomy, and sandbox support, but generally assume that the target repository can be evaluated in a prepared execution environment with dependencies installed and a runnable test suite.
They therefore focus on modifying source code rather than configuring the execution environment itself;
\textbf{(2) Repository Setup.}
Recent work has begun to study this setup phase across test-execution, Dockerized repository setup, and research-repository reproduction settings, including ExecutionAgent~\cite{executionagent}, Repo2Run~\cite{hu2026reporun}, and SUPER~\cite{bogin2024super}.
However, existing setup methods and evaluations still treat each repository or task as an isolated episode, so a fix discovered for one project cannot directly inform future repositories.


\noindent \underline{$\blacktriangleright$ \textbf{Experience and Skill Reuse in LLM Agents.}}
Prior work has explored several forms of reusable agent experience. 
Executable skill libraries, such as Voyager~\cite{wang2023voyager} and SkillCraft~\cite{chen2026skillcraft}, store reusable programs or tool-use routines that can be retrieved and composed in later tasks.
Natural-language memory systems, such as Reflexion~\cite{reflexion} and ExpeL~\cite{zhao2024expel}, instead distill prior trajectories into verbal feedback or general insights.
More recent systems, including Agent-KB~\cite{agentkb2025}, EvolveR~\cite{wu2025evolver}, and Mem\textsuperscript{p}~\cite{memp}, add richer organization, refinement, and memory-maintenance mechanisms.
These approaches show the value of experience reuse, but they are typically organized around general task workflows rather than setup-specific diagnostic signals.



\noindent \underline{$\blacktriangleright$ \textbf{Verification and Quality Assurance.}}
Existing setup verification commonly relies on either proxy-based or execution-based signals.
EnvBench~\cite{envbench} uses missing-import analysis for Python repositories and compilation success for JVM repositories. ExecutionAgent~\cite{executionagent} and Repo2Run~\cite{hu2026reporun} instead validate setup by running build or test commands in the configured environment. These signals provide useful executable evidence, but they often collapse heterogeneous causes (e.g., setup defects, optional dependencies, missing external service data) into a single pass/fail outcome.
LLM-as-judge and multi-agent evaluation frameworks, such as CourtEval~\cite{courteval} and Agent-as-a-Judge~\cite{agentasajudge}, introduce role separation or agentic evaluation for assessing LLM outputs subjectively.
\oursys{} adapts role separation to environment setup, where claims can be checked through targeted container commands, enabling responsibility-sensitive verification. 

\vspace{-.25cm}
\section{Conclusion}
\label{sec:conclusion}
\vspace{-.25cm}

We introduced \oursys, an experiential learning agent for
\ourproblem that combines three mechanisms:
\emph{speculative execution} for safe trial under non-invertible state;
an \xpudb that turns past failures into self-curating
knowledge; and an in-loop \emph{Verifier} sub-agent that provides
diagnostic feedback during setup. For unbiased outcome assessment,
we further design a structurally independent \emph{Prosecutor--Judge}
protocol that decouples investigation from adjudication. Across 100 Python
repositories, \oursys substantially outperforms both general-purpose
and specialized baselines, turning environment setup from a brittle,
one-shot bottleneck into an experience-accumulating, independently
verifiable component of the LLM-agent stack. 
Nevertheless, current evaluation focuses on single-run containerized setup, and future work will extend \oursys to broader language ecosystems, enable cost-efficient repeated evaluation, and scale telemetry-driven experience evolution for long-term validation.



\newpage

\appendix
\appendix

\section{XPU and Retrieval Implementation}
\label{app:xpu-details}

\subsection{Example XPU}

\begin{tcolorbox}[
  enhanced,
  breakable,
  colback=gray!3,
  colframe=gray!45,
  coltitle=black,
  title={Example XPU Entry},
  fonttitle=\bfseries,
  boxrule=0.4pt,
  arc=1.2mm,
  left=1mm,
  right=1mm,
  top=1mm,
  bottom=1mm
]
\begin{PromptVerbatim}
{
  "id": "xpu_poetry_lock_conflict",
  "signals": {
    "keywords": ["poetry.lock", "pyproject.toml", "dependency conflict"],
    "regex": ["Because .* depends on .*", "version solving failed"],
    "situation_triggers": [
      "Poetry-managed project where manual pip fallback risks losing the lock graph"
    ]
  },
  "advice_nl": [
    "Do not bypass Poetry with manual pip installation.",
    "Preserve the locked dependency graph and resolve the conflict inside Poetry.",
    "Run Poetry installation end-to-end after checking the lock file."
  ],
  "atoms": [
    {"name": "inspect_file", "args": {"path": "pyproject.toml"}},
    {"name": "inspect_file", "args": {"path": "poetry.lock"}},
    {"name": "shell", "args": {"cmd": "poetry install --no-interaction"}}
  ],
  "telemetry": {
    "hits": 63,
    "successes": 37,
    "failures": 15
  }
}
\end{PromptVerbatim}
\end{tcolorbox}

\subsection{Anchor-Based Retrospective Evaluation Protocol}
\label{app:delayed-audit}

Whenever the Retriever Agent is invoked, it first audits the outcome of the previous retrieval. 
For each previously recommended \xpu, the Retriever Agent assigns a categorical verdict
\(v \in \{\textit{success}, \textit{failure}, \textit{neutral}\}\) based on the subsequent setup trajectory.

\begin{enumerate}[leftmargin=*, nosep]
    \item \textbf{Anchor.} At retrieval time, record the current trajectory length and the identifiers of the recommended \xpu entries, establishing a temporal reference point.

    \item \textbf{Extract.} On the next retrieval call, extract up to five subsequent steps from the main agent's trajectory after the anchor point.

    \item \textbf{Judge.} An LLM judge determines whether each recommended \xpu contributed to resolving the observed problem. The verdict is \textit{success} if the advice was adopted and the problem was resolved or improved; \textit{failure} if the advice was adopted but the problem persisted or worsened; and \textit{neutral} if adoption or causal contribution cannot be determined.

    \item \textbf{Update.} The telemetry counters of each recommended \xpu are updated atomically according to the verdict:
    \begin{equation}
        (\textit{successes},\, \textit{failures}) \mathrel{+}=
        \begin{cases}
            (1, 0), & \text{if } v = \textit{success}, \\
            (0, 1), & \text{if } v = \textit{failure}, \\
            (0, 0), & \text{if } v = \textit{neutral}.
        \end{cases}
        \label{eq:audit_update}
    \end{equation}
\end{enumerate}

\section{Agent Prompt Excerpts}
\label{app:agent-prompts}

This appendix provides selected excerpts from the role-specific prompts used by SETUPX. We include only the parts that define each agent's role, action boundary, and output contract. The complete prompts will be released with the code artifact.

\begin{tcolorbox}[enhanced,breakable,colback=gray!3,colframe=gray!45,coltitle=black,title={Setup Agent Prompt Excerpt},fonttitle=\bfseries,boxrule=0.4pt,arc=1.2mm,left=1mm,right=1mm,top=1mm,bottom=1mm]
\begin{PromptVerbatim}
You are an expert DevOps agent tailored for environment setup.
You have access to a Linux terminal and an external eXPerience Unit (XPU).

## Action Types — Purpose and When to Use

### SHELL_COMMAND
Execute any shell command directly in the container.

### TRY_XPU_SUGGESTION
Apply a proven fix from the XPU knowledge base inside a snapshot sandbox.

### SET_ENV
Persist an environment variable across all subsequent commands.

### ROLLBACK_ENV
Pop any number of frames off the snapshot stack and return the container to
any earlier known-good checkpoint, not only the latest one.

### VERIFY
Trigger the full pytest verification pipeline.

### FINISH
Signal that the task is complete. ONLY call after a successful VERIFY.

You MUST respond in JSON format with this schema:
{
  "thought": "Analyze the current state and the cause of the error, then explain why you chose this action...",
  "action_type": "SHELL_COMMAND | TRY_XPU_SUGGESTION | SET_ENV | ROLLBACK_ENV | VERIFY | FINISH",
  "content": {
    // For SHELL_COMMAND:
    "command": "pip install numpy",

    // For TRY_XPU_SUGGESTION:
    "xpu_suggestion_id": "suggestion_123",
    "command": "pip install numpy==1.23.5",
    "reasoning": "The XPU suggests downgrading numpy; this matches the error closely."

    // For SET_ENV:
    "env_key": "VAR_NAME",
    "env_value": "value"

    // For ROLLBACK_ENV:
    // "n_frames": 1   // default 1; pass >=2 to go back to an earlier checkpoint.
    //
    // For VERIFY / FINISH:
    // (VERIFY needs no extra fields)
    // FINISH requires: "message": "environment setup complete"
  }
}
\end{PromptVerbatim}
\end{tcolorbox}

\begin{tcolorbox}[enhanced,breakable,colback=gray!3,colframe=gray!45,coltitle=black,title={Retriever Agent Prompt Excerpts},fonttitle=\bfseries,boxrule=0.4pt,arc=1.2mm,left=1mm,right=1mm,top=1mm,bottom=1mm]
\begin{PromptVerbatim}
You are an XPU experience-retrieval assistant. Given a list of candidate XPU experiences, your task is to pick the Top-K that best match the current deployment situation.

## Selection rules
1. Exact match first: XPUs whose advice_nl directly addresses the current problem rank highest.
2. Telemetry as reference: pay attention to each XPU's historical hit / success / failure counts, but do not discard one merely because it has many failures — judge whether the previous failure scenarios are similar to the current one. If they are not, the XPU may still be effective.
3. Drop the irrelevant: if an XPU's advice is completely unrelated to the current problem, do not pick it.
4. Pick at most {k}.

%## Selection rules
%1. Exact match first: XPUs whose advice_nl directly solves the current problem rank first.
%2. Telemetry as reference: pay attention to each XPU's historical hits / successes / failures, but do not exclude an XPU just because it has many failures.
%3. Exclude irrelevant items outright: if an XPU's advice has nothing to do with the current problem, do not select it.
%4. Select at most {k} items.

You are an XPU experience-audit assistant. Your task: judge whether the XPU experiences recommended to the Agent last time helped the deployment.

%## Decision rules
%- success: the Agent adopted the XPU's idea, and subsequent steps show the problem was solved or clearly improved.
%- failure: the Agent adopted the XPU's idea, but the problem was not solved or new problems were introduced.
%- neutral: cannot tell whether the Agent adopted the suggestion, or the suggestion is unrelated to the subsequent steps.

## Decision rules (judge each XPU separately)
- success: the Agent adopted the XPU's idea (possibly with a different command but the same approach), and subsequent steps show the problem was solved or clearly improved.
- failure: the Agent adopted the XPU's idea, but the problem was not solved or new problems were introduced.
- neutral: cannot tell whether the Agent adopted the suggestion, or the suggestion is unrelated to the subsequent steps.

\end{PromptVerbatim}
\end{tcolorbox}

\begin{tcolorbox}[enhanced,breakable,colback=gray!3,colframe=gray!45,coltitle=black,title={In-loop Verifier Agent Prompt Excerpt},fonttitle=\bfseries,boxrule=0.4pt,arc=1.2mm,left=1mm,right=1mm,top=1mm,bottom=1mm]
\begin{PromptVerbatim}
You are a verification Agent working inside a Docker container.
Your task: examine whether the environment configured by the Setup Agent is acceptable, and report the result truthfully.

You are not in charge of fixing anything; you only run tests, observe results, and pass judgment.

## Verification procedure
1. Structure reconnaissance: `ls` the project root, locate pyproject.toml / setup.cfg / pytest.ini / tox.ini, etc.
2. Locate the test suite: confirm the test directory and framework. (pytest / unittest / tox, ...).
3. Run the tests in the project's native way and collect results.
4. Analyze failure causes and make a judgment.
5. If the project has no tests at all, write a smoke test under /tmp/ to verify basic environment usability.

## Hard constraints (violation invalidates the verdict)
- Install no packages.
- Modify no environment configuration.
- Modify no file under /workspace/repo.
- write_file may only write into /tmp/.
\end{PromptVerbatim}
\end{tcolorbox}

\begin{tcolorbox}[enhanced,breakable,colback=gray!3,colframe=gray!45,coltitle=black,title={Prosecutor and Judge Prompt Excerpts},fonttitle=\bfseries,boxrule=0.4pt,arc=1.2mm,left=1mm,right=1mm,top=1mm,bottom=1mm]
\begin{PromptVerbatim}
You are the prosecutor. Your stance is skeptical: both the Setup Agent and the Verifier may make mistakes, take shortcuts, or deceive themselves.
You must verify independently using actual evidence inside the container, not trust their self-reports.

## Mandatory investigation procedure (in order, no skipping)
Step 0: identify the project language and build tool.
Step 1: verify that core dependencies are available.
Step 2: exercise the entry commands declared in README.
Step 3: run the test suite yourself.
Step 4: adjudicate each failure category.
Step 5: cross-check the credibility of the Verifier's conclusion.

You are the Judge. Your job is to verify the prosecutor's charges one by one and then issue a verdict.

You are not the prosecutor — you do not conduct open-ended investigation. Your responsibility: for each charge raised by the prosecutor, execute 1–2 verification commands to confirm whether that charge holds.

## Final verdict
- >=1 charge upheld after your verification -> guilty
- All charges dismissed -> not_guilty
\end{PromptVerbatim}
\end{tcolorbox}

\begin{tcolorbox}[enhanced,breakable,colback=gray!3,colframe=gray!45,coltitle=black,title={XPU Distiller Prompt Excerpt},fonttitle=\bfseries,boxrule=0.4pt,arc=1.2mm,left=1mm,right=1mm,top=1mm,bottom=1mm]
\begin{PromptVerbatim}
You are a senior expert in Python project environment configuration and dependency issues.
You will be given a complete agent trajectory (executed commands and error logs) from an automated environment setup of a repository, and (when available) the Phase 2 Prosecutor-Judge adjudication signal phase2_context for that trajectory.

Mandatory four-step distillation procedure:
[Step 1: Verdict-Aware Ingestion] Eat the verdict first, then read the trajectory.
Step 2: Forward Attribution] Trace each problem forward to the action that actually fixed it.
[Step 3: Schema-Level Distillation] Map each problem-fix pair onto the XPU schema.

%Your task:
%1. Carefully analyze the entire trajectory and identify all independent environment problems.
%2. For each independent problem, judge whether it is worth distilling into a reusable environment experience (XPU).
%3. For each problem worth distilling, generate one structured XPU entry.

Distillation principles:
- prosecution_charges are the cleanest causal knowledge source — distill from them first.
- Even when verdict=guilty, distill the generalizable patterns within.
- One XPU = one root cause; never mix several unrelated problems.
- Do not produce an id field; the system assigns a unique ID automatically.
\end{PromptVerbatim}
\end{tcolorbox}

\section{Action Space}
\label{sec:action_space}
The agent executes the selected action from the action space $\mathcal{A}_{\text{act}} = \{a_1, \ldots, a_6\}$:

\begin{itemize}[leftmargin=*, nosep]
    \item \texttt{SHELL\_COMMAND}: Execute an arbitrary shell command---the primary action for diagnosis and repair.
    \item \texttt{TRY\_XPU\_SUGGESTION}: Speculatively apply a retrieved \xpu entry under the snapshot protection described in Section~\ref{sec:speculative}.
    \item \texttt{SET\_ENV}: Persistently set an environment variable in the container.
    \item \texttt{ROLLBACK\_ENV}: Roll back the container to the most recent snapshot on the LIFO stack, or to a specific earlier checkpoint.
    \item \texttt{VERIFY}: Invoke an in-loop Verifier Agent (Section~\ref{sec:inloop_verifier}) to assess environment readiness.
    \item \texttt{FINISH}: Declare setup complete and proceed to Phase~2 Prosecutor-Judge verification.
\end{itemize}

\subsection{\texttt{TRY\_XPU\_SUGGESTION}}
\label{sec:try_xpu}
The speculative execution protocol proceeds as follows:

\begin{enumerate}[leftmargin=*, nosep]
    \item \textbf{Checkpoint}: Snapshot the current container via \texttt{docker commit}, pushing the image onto the LIFO stack $\mathcal{S}$.
    \item \textbf{Adapt}: The agent reads the \xpu's \textit{advice\_nl}, leverages its full conversation context (recent history, observed versions, repository structure), and generates concrete commands tailored to the current repository. If the agent produces no command, the system falls back to rendering the \xpu's \textit{atoms} via a type-aware \emph{Atom Rendering Engine} that maps 12 predefined atom types (e.g., \texttt{pip\_install}, \texttt{apt\_install}) to executable bash commands.
    \item \textbf{Trial}: Execute the adapted commands sequentially; halt on any non-zero exit code.
    \item \textbf{Verify}: Compare the error state before and after execution to determine the trial outcome.
    \item \textbf{Decision}: Retain changes on success trial; roll back to the snapshot otherwise.
\end{enumerate}

\section{Multi-repository Track Evaluation Rubric}
\addcontentsline{toc}{section}{Appendix A: Multi-repository Track Evaluation Rubric}
\label{app:rubric}

\subsection*{D.1 Overview}

For the 22 non-atomic families, a verdict is assigned by direct
in-container inspection rather than by the prosecutor--judge protocol
used on the single-repository benchmark, because family-level
usability spans multiple containers, services, and configuration
consistency that no single adjudication protocol covers.

Each family is independently rated by three evaluators using the
rubric below. Disagreements are resolved through discussion until
consensus is reached.

\subsection*{D.2 Five-Check Rubric}

For each family's host container, the following five checks are
performed:

\paragraph{C1. Clone integrity.} The host repository is fully cloned
and contains all expected source files. (Verified via
\texttt{ls /workspace/repo} and \texttt{git status}.)

\paragraph{C2. Editable install.} The host package is installed in
editable mode and resolves correctly. (Verified via
\texttt{pip list -e}.)

\paragraph{C3. Module / CLI availability.} The primary module of the
host package is importable, and the CLI entry point (if any) responds
to \texttt{--version} or \texttt{--help}. (Verified via
\texttt{python -c "import <module>"} and \texttt{<cli> --version}.)

\paragraph{C4. Pytest collection.} The host repository's test suite
can be collected by \texttt{pytest --co} without fatal errors. The
number of collected tests is recorded.

\paragraph{C5. Dependency consistency.} \texttt{pip check} reports no
broken or inconsistent dependencies in the venv.

For families involving cross-container protocol layers (sibling
services, Kubernetes deployments, plugin entry-point registration),
additional protocol checks are performed inside the relevant
container, e.g., service connectivity via socket connection,
\texttt{kubectl get pods} for Kubernetes families, or
\texttt{importlib.metadata.entry\_points()} for plugin-host families.

\subsection*{D.3 Four-Tier Verdict}

Verdicts are assigned based on the worst-evidence $+$ overall
main-path usability principle:

\begin{center}
\small
\begin{tabular}{@{}l p{10.5cm}@{}}
\toprule
\textbf{Verdict} & \textbf{Criteria} \\
\midrule
\textbf{Full}    & All five checks pass; protocol layer (if any) is fully realized: sibling services running, Kubernetes pods Up, plugin entry points registered. \\
\addlinespace
\textbf{Mostly}  & Host setup is complete and the main path is functional, but with one or two bounded minor gaps (e.g., a few collection errors, one sibling missing, an unfinished helm step). \\
\addlinespace
\textbf{Partial} & Host installation is partial; some core dependencies missing or main module imports with errors. The intended workflow does not run end-to-end. \\
\addlinespace
\textbf{Shallow} & Host repository is cloned but core dependencies are not installed; \texttt{import} fails immediately, or \texttt{git clone} itself failed. \\
\bottomrule
\end{tabular}
\end{center}

The \textbf{Full--Mostly boundary} is determined by whether one
independently fixable minor gap exists. The \textbf{Shallow} tier
qualitatively differs from the other three: it represents ``did not
actually install'' rather than ``installed with limitations.''

Pytest collection serves as a key signal: successful collection of
the majority of tests indicates that host setup is substantively
complete.

\section{Per-family Verdicts of multi-repository}
\label{app:per-family-verdicts}

This appendix reports per-family verdicts for the non-atomic
ecosystems used in our multi-repository experiment. For each family we run
two independent setup pipelines---\textsc{SetupX} and the \textsc{Qwen
Code} baseline---and rate the resulting container against the five
checks of Appendix~\ref{app:rubric} (clone integrity, editable install,
module/CLI availability, \texttt{pytest~--co}, and \texttt{pip check})
plus protocol-layer checks where relevant (sibling-service ports,
Kubernetes pods, plugin entry-point registration). Verdicts use the
four tiers defined there: \textbf{Full}, \textbf{Mostly},
\textbf{Partial}, \textbf{Shallow}. The \emph{primary repo} column
gives the entry repository in the family's selection record; the
\emph{Pattern} column summarises the cross-repository dependency type;
the \emph{Notes} column gives a one-line contrast of the two
pipelines' behaviour on that family.

{\scriptsize
\setlength{\tabcolsep}{3pt}
\renewcommand{\arraystretch}{1.08}

\begin{longtable}{@{}r p{3.7cm} p{2.05cm} c c@{}}
\caption{Per-family verdicts on the 22 non-atomic ecosystems. \textsc{SetupX} reaches Full on 6/22 and
Mostly on 11/22; \textsc{Qwen Code} reaches Full on 1/22 and Mostly on
16/22. Both systems clear the $\geq$Mostly bar on 17/22, but the
internal composition of the verdicts differs (see
Appendix~\ref{app:multi-repo_cases} for a worked example).}
\label{tab:per-family-verdicts} \\

\toprule
\# & Family (primary repo) & Pattern & \textsc{SetupX} & \textsc{Qwen} \\
\midrule
\endfirsthead

\multicolumn{5}{l}{\scriptsize\itshape Table~\ref{tab:per-family-verdicts} -- continued} \\
\toprule
\# & Family (primary repo) & Pattern & \textsc{SetupX} & \textsc{Qwen} \\
\midrule
\endhead

\midrule
\multicolumn{5}{r}{\scriptsize\itshape continued on next page} \\
\endfoot

\bottomrule
\endlastfoot

1 & \repo{terminusdb-client-python} & client/server & Full & Mostly \\
\multicolumn{5}{@{}p{0.96\textwidth}@{}}{\emph{Notes:} \textsc{SetupX} runs 122 integration tests inside setup (sibling \texttt{:6363} server up); \textsc{Qwen} reaches port connectivity but stops at test collection.} \\[2pt]

2 & \repo{ansible/awx-operator} & K8s deployment & Mostly & Mostly \\
\multicolumn{5}{@{}p{0.96\textwidth}@{}}{\emph{Notes:} Both bring all AWX pods (web/task/postgres) Up on a KIND cluster; \textsc{Qwen} ships no editable Python install for any package.} \\[2pt]

3 & \repo{HumanSignal/label-studio} & full platform & Shallow & Partial \\
\multicolumn{5}{@{}p{0.96\textwidth}@{}}{\emph{Notes:} \textsc{SetupX} hits a 5\,min \texttt{git clone} timeout on the 1\,GB monorepo, leaving an empty checkout; \textsc{Qwen} clones 81 packages but the host package is not editable and the CLI is absent.} \\[2pt]

4 & \repo{scrapy/scrapy} & deploy + plugin host & Mostly & Mostly \\
\multicolumn{5}{@{}p{0.96\textwidth}@{}}{\emph{Notes:} \textsc{SetupX} covers the host only (4 siblings missing); \textsc{Qwen} installs \texttt{scrapyd}/\texttt{splash}/\texttt{redis}/\texttt{scrapy-redis}, but \texttt{pytest} is absent and the Scrapyd process is a zombie.} \\[2pt]

5 & \repo{saleor/saleor} & Django commerce & Mostly & Mostly \\
\multicolumn{5}{@{}p{0.96\textwidth}@{}}{\emph{Notes:} Both collect 17{,}228 tests with no import errors; in both, the \texttt{saleor} host package itself is not truly editable and \texttt{import saleor} fails.} \\[2pt]

6 & \repo{odoo/odoo} & framework + OCA addons & Mostly & Mostly \\
\multicolumn{5}{@{}p{0.96\textwidth}@{}}{\emph{Notes:} \textsc{SetupX} runs 1{,}245 base tests but ships no OCA siblings; \textsc{Qwen} clones the three OCA repos and keeps \texttt{odoo-bin} alive on \texttt{:8069}, but no editable install.} \\[2pt]

7 & \repo{intake/intake} & plugin host & Mostly & Mostly \\
\multicolumn{5}{@{}p{0.96\textwidth}@{}}{\emph{Notes:} \textsc{SetupX} ships core editable + 9 driver registrations; \textsc{Qwen} adds \texttt{intake-xarray}/\texttt{sql}/\texttt{geopandas} as editable siblings (794 tests collectable across packages).} \\[2pt]

8 & \repo{robotframework/robotframework} & plugin host & Mostly & Full \\
\multicolumn{5}{@{}p{0.96\textwidth}@{}}{\emph{Notes:} \textsc{Qwen} adds \texttt{SeleniumLibrary}/\texttt{RequestsLibrary}/\texttt{Browser} siblings + 2{,}318 utests OK; \textsc{SetupX} keeps the host alone, none of the three siblings installed.} \\[2pt]

9 & \repo{napari/napari} & Qt plugin host & Full & Mostly \\
\multicolumn{5}{@{}p{0.96\textwidth}@{}}{\emph{Notes:} Both instantiate a headless \texttt{Viewer} under \texttt{QT\_QPA\_PLATFORM=offscreen}; \textsc{Qwen} reports one dev-version warning and 3 collection errors.} \\[2pt]

10 & \repo{apache/airflow} & client/server + deploy & Mostly & Shallow \\
\multicolumn{5}{@{}p{0.96\textwidth}@{}}{\emph{Notes:} \textsc{SetupX} ships \texttt{core}+\texttt{client}+\texttt{cncf-kubernetes provider} all editable, both Postgres/Redis sibling ports up; \textsc{Qwen} has only the client editable, \texttt{import airflow} fails.} \\[2pt]

11 & \repo{matrix-org/synapse} & multi-service homeserver & Mostly & Mostly \\
\multicolumn{5}{@{}p{0.96\textwidth}@{}}{\emph{Notes:} \textsc{SetupX} switches to a Poetry venv and rebuilds the Rust extension on aarch64; \textsc{Qwen} runs the homeserver process plus \texttt{matrix-client}/\texttt{sydent} sibling editables.} \\[2pt]

12 & \repo{ckan/ckan} & runtime + plugin host & Full & Mostly \\
\multicolumn{5}{@{}p{0.96\textwidth}@{}}{\emph{Notes:} \textsc{SetupX} collects 3{,}358 tests and runs alembic Upgrade DB to completion; \textsc{Qwen} keeps CKAN listening on \texttt{:5000} but \texttt{pytest} is missing and Solr \texttt{:8983} is unreachable.} \\[2pt]

13 & \repo{flyteorg/flyte} & Go backend + Py SDK & Partial & Mostly \\
\multicolumn{5}{@{}p{0.96\textwidth}@{}}{\emph{Notes:} \textsc{SetupX} picks the wrong language path (Python image vs.\ Go monorepo) and crashes on embedded-postgres root pollution; \textsc{Qwen} ships Flyte~2 SDK + \texttt{flyteidl2} as editable.} \\[2pt]

14 & \repo{MISP/MISP} & PHP server + Py SDK & Partial & Partial \\
\multicolumn{5}{@{}p{0.96\textwidth}@{}}{\emph{Notes:} \textsc{SetupX} skips the PHP server entirely (Python base image), only \texttt{PyMISP} is in place; \textsc{Qwen} finds \texttt{:80} occupied by a Saleor leftover and \texttt{:443} refused; \texttt{misp-modules} cloned only.} \\[2pt]

15 & \repo{bitcart/bitcart} & multi-repo workflow & Partial & Mostly \\
\multicolumn{5}{@{}p{0.96\textwidth}@{}}{\emph{Notes:} \textsc{SetupX} hits the Python~3.10 vs.\ \texttt{>=3.12} floor; \texttt{:5432}/\texttt{:6379} both refuse. \textsc{Qwen} uses \texttt{uv} + Python~3.14.4 and brings 4 sibling containers + Backend \texttt{:8000} up.} \\[2pt]

16 & \repo{bottlecapdave/homeassistant-octopusenergy} & HA custom integration & Mostly & Partial \\
\multicolumn{5}{@{}p{0.96\textwidth}@{}}{\emph{Notes:} \textsc{SetupX} \texttt{apt}-installs \texttt{python3.13-venv} matching the manifest pin (759 tests, 0 error); \textsc{Qwen}'s venv pins HA \texttt{2024.3.3} against a \texttt{2025.11+} requirement, yielding 96 collection errors.} \\[2pt]

17 & \repo{frappe/erpnext} & framework + app & Partial & Shallow \\
\multicolumn{5}{@{}p{0.96\textwidth}@{}}{\emph{Notes:} \textsc{SetupX} pip-installs \texttt{frappe} + editable \texttt{erpnext} but ships no \texttt{bench}/MariaDB; \textsc{Qwen}'s Python~3.11 is rejected by source-level PEP~695 syntax (\texttt{type ConfType = ...}).} \\[2pt]

18 & \repo{internetarchive/brozzler} & archiving workflow & Full & Mostly \\
\multicolumn{5}{@{}p{0.96\textwidth}@{}}{\emph{Notes:} Both install \texttt{brozzler}+\texttt{warcprox} as editables and bring up a RethinkDB sibling on \texttt{:28015}; \textsc{SetupX} additionally installs Chromium 147 with the full sandbox stack.} \\[2pt]

19 & \repo{jazzband/wagtailmenus} & Wagtail plugin & Mostly & Mostly \\
\multicolumn{5}{@{}p{0.96\textwidth}@{}}{\emph{Notes:} Both run \texttt{runtests.py} cleanly with all 175 tests OK; \textsc{Qwen} additionally installs \texttt{wagtail-localize}, \texttt{-localize-git}, and \texttt{-grapple} as editable siblings.} \\[2pt]

20 & \repo{jupyterhub/jupyterhub} & Hub plugin host & Full & Mostly \\
\multicolumn{5}{@{}p{0.96\textwidth}@{}}{\emph{Notes:} \textsc{SetupX} ships \texttt{oauthenticator} editable with 11 OAuth-provider entry points registered and a full \texttt{pytest} pass; \textsc{Qwen} keeps Hub running and adds \texttt{kubespawner}/\texttt{dockerspawner}/\texttt{batchspawner}.} \\[2pt]

21 & \repo{mov-cli/mov-cli} & CLI plugin host & Mostly & Mostly \\
\multicolumn{5}{@{}p{0.96\textwidth}@{}}{\emph{Notes:} Both load three plugins via the \texttt{mov-cli-*} module-naming convention; \textsc{SetupX} additionally executes a live YouTube search returning 13 hits dated 2026-05.} \\[2pt]

22 & \repo{qiboteam/qibolab} & quantum backend driver & Full & Mostly \\
\multicolumn{5}{@{}p{0.96\textwidth}@{}}{\emph{Notes:} \textsc{SetupX} instantiates \texttt{create\_platform("dummy")} into a 5-qubit \texttt{Platform} with the qibolab backend entry point registered; \textsc{Qwen} ships no \texttt{pytest}, no CLI, with the host repo at \texttt{/workspace} not the expected \texttt{/workspace/repo}.} \\[2pt]

\end{longtable}
}

\paragraph{Aggregate.}
\textsc{SetupX}'s distribution is Full~$=6$, Mostly~$=11$,
Partial~$=4$, Shallow~$=1$, with $\geq$Mostly on $17/22$.
\textsc{Qwen Code}'s distribution is Full~$=1$, Mostly~$=16$,
Partial~$=3$, Shallow~$=2$, also with $\geq$Mostly on $17/22$. The two
systems tie on the $\geq$Mostly bar but their verdict structure
differs: \textsc{SetupX}'s six \emph{Full} verdicts are concentrated on
families where the protocol layer is end-to-end observable
(real sibling containers, real protocol handshakes, real integration
tests), while sixteen of \textsc{Qwen Code}'s \emph{Mostly} verdicts
sit at ``clone + install + collect'' without ever crossing into the
cross-container protocol layer (Appendix~\ref{app:multi-repo_cases}
develops the \texttt{terminusdb-client-python} family in detail as a
worked example of this asymmetry).

\paragraph{Failure modes.}
\textsc{SetupX}'s four Partial/Shallow cases concentrate on (i)~the
5-minute \texttt{git clone} ceiling failing on $\sim$1\,GB monorepos
(\texttt{label-studio}); (ii)~cross-language families colliding with
the default Python base image (\texttt{flyte} vs.\ Go,
\texttt{MISP} vs.\ PHP, \texttt{frappe} via the wrong strategy); and
(iii)~Python-version floors above the base image
(\texttt{bitcart} 3.10 vs.\ 3.12+ requirement).
\textsc{Qwen Code}'s five Partial/Shallow cases come predominantly
from (i)~missing editable install of the host package leading to
\texttt{import} failure (\texttt{airflow}, \texttt{frappe});
(ii)~mis-pinned host versions
(\texttt{homeassistant-octopusenergy} pinned to HA~2024.3 against a
2025.11+ requirement); and (iii)~port hijacking by leftover
containers from a previous family
(\texttt{MISP}'s \texttt{:80} occupied by a stale Saleor process).
The two failure-mode groups are largely disjoint, indicating that
the bottleneck of multi-repository setup is split between
\emph{protocol adaptation} (\textsc{SetupX}-side) and
\emph{dependency-assembly semantics} (\textsc{Qwen}-side) at
different layers.

\section{Representative Multi-repository Cases}
\label{app:multi-repo_cases}

We zoom into two of the 22 family verdicts in
Appendix~\ref{app:per-family-verdicts} as worked examples. The two
cases are chosen by structural criteria rather than by outcome:
\textbf{terminusdb-client-python} is the only pure client/server
family in the 22 (its cross-repository dependency is a single HTTP
protocol against a sibling server, with no plugin-host or
multi-service confounders), and \textbf{frappe/erpnext} is one of the
families on which \textsc{SetupX+XPU} itself underperforms (Partial),
chosen so that the case pairing covers both ends of the verdict scale
rather than only families where SetupX wins.

\subsubsection*{terminusdb-client-python (client/server, SetupX+XPU=Full / Qwen Code=Mostly)}

\textbf{Run identifiers.} \textsc{SetupX+XPU} on container
\texttt{bfd7bd0e6b8f}; \textsc{Qwen Code} on container
\texttt{4462896bd3c7}.

terminusdb-client-python is the Python SDK for TerminusDB. Its integration tests
(e.g., \texttt{Client.connect}, \texttt{db\_create} / \texttt{db\_delete}) all require
sending HTTP requests to a running TerminusDB server and parsing the response.
A standalone \texttt{pip install -e .} only covers unit tests over the client's
internal data structures (schema / document / WOQL AST construction); the full
integration suite requires a TerminusDB server running in parallel.

\paragraph{SetupX+XPU trajectory (17 steps).} The agent inspects
\texttt{pyproject.toml} and \texttt{conftest.py}, identifies that \texttt{integration\_tests/}
target \texttt{localhost:6363}, then runs \texttt{apt install docker.io} followed by
\texttt{docker pull/run terminusdb-server} to bring up the sibling. \texttt{curl /api/info}
confirms HTTP 200. After installing the client and test dependencies,
\texttt{pytest integration\_tests/test\_client.py} reports 122 passed / 5 skipped,
and the agent halts on its own assessment that setup is complete.

\paragraph{Qwen Code trajectory (setup completed).} Qwen Code likewise recognizes
the client/server protocol, brings up \texttt{terminusdb-server} on port 6363, and
performs an editable install of the client SDK in a venv. \texttt{pytest --co}
collects 1499 tests and \texttt{pip check} reports clean. However, the setup
terminates at the ``tests can be collected'' stage without invoking the integration
suite. Separately, the container's default \texttt{PATH} is not linked to
\texttt{/workspace/.venv/bin/}, so bare \texttt{python} / \texttt{pytest} commands
return \texttt{executable not found} and the full venv path must be used explicitly.

\paragraph{Verdict justification.} Evaluators independently apply the five-check
rubric and the protocol-layer extensions (Appendix~\ref{app:rubric}) to both containers:

\begin{center}
\small
\begin{tabular}{lcc}
\toprule
Check & SetupX+XPU & Qwen Code \\
\midrule
Clone integrity & PASS & PASS \\
Editable install (host package in \texttt{pip list -e}) & PASS & PASS \\
\texttt{pip check} clean & PASS & PASS \\
Protocol: sibling server up + HTTP handshake & PASS & PASS \\
Protocol: integration tests invoked during setup & 122 passed & \textbf{not invoked} \\
Module import (default \texttt{PATH}) & PASS & FAIL (requires full venv path) \\
\bottomrule
\end{tabular}
\end{center}

Both runs satisfy the host-side five-check rubric and the basic protocol layer
(sibling server running, port reachable). The decisive gap is the last
protocol-layer check: Appendix~\ref{app:rubric} treats end-to-end protocol-layer
verification (here, executing the integration suite that exercises the
client/server handshake) as a verdict criterion rather than an optional
follow-up. \textsc{SetupX+XPU} runs the suite during setup and observes 122
passes; \textsc{Qwen Code} stops at \texttt{pytest --co} without ever invoking the
suite. A secondary, cosmetic demerit is that Qwen Code's default \texttt{PATH} is
not linked to its venv, requiring downstream callers to use the full venv path.
Under the rubric, \textsc{SetupX+XPU} is rated \textbf{Full} (all five checks plus
the protocol-layer end-to-end check pass) and \textsc{Qwen Code} is rated
\textbf{Mostly} (host setup complete and main path functional, with a bounded gap
on the end-to-end check and the \texttt{PATH} cosmetic issue).

\subsubsection*{frappe/erpnext (multi-service deployment, SetupX+XPU=Partial / Qwen Code=Shallow)}

\textbf{Run identifiers.} \textsc{SetupX+XPU} on container
\texttt{3bc8bfb590f8}; \textsc{Qwen Code} on container
\texttt{7b071e5b5237}.

frappe/erpnext is an ERP business application. Its standard deployment relies on
the \texttt{bench} toolchain (\texttt{bench init} / \texttt{bench install-app})
to host multiple sibling apps, with parallel MariaDB (port 3306) and Redis
siblings providing persistence and cache. The latest source's \texttt{pyproject.toml}
requires Python \texttt{>=3.14,<3.15} (the source already uses PEP~695 \texttt{type}
aliases such as \texttt{type ConfType = ...}); the container's default base image
ships Python 3.11.

\paragraph{SetupX+XPU trajectory (53 steps).} The agent inspects the source and
identifies the Python version constraint, then takes a \emph{sidestep} strategy
rather than satisfying the constraint directly: it runs \texttt{apt install
python3.13} and creates a Python 3.13 venv at \texttt{/workspace/venv}, then installs
the released v15 tag via \texttt{pip install
git+https://github.com/frappe/frappe@version-15} (non-editable, frappe 15.107.0; the
v15 release has a lower Python floor than the latest dev source) and installs
erpnext editable from \texttt{/workspace/repo} (15.106.0). Both \texttt{import frappe}
and \texttt{import erpnext} succeed. \texttt{pytest --co} lists 260 tests but reports
171 collection errors (missing frappe site context). After creating a placeholder
\texttt{sites/test\_site/site\_config.json}, the agent attempts \texttt{apt install
docker.io} to start a MariaDB sibling; the MariaDB sibling does not come up, the
bench CLI is never installed, and the bench toolchain pipeline is not invoked.

\paragraph{Qwen Code trajectory.} Qwen Code uses the container's default Python
3.11.2 without upgrading. \texttt{pip install -e /workspace} immediately raises a
\texttt{SyntaxError} on the source's PEP~695 \texttt{type} alias, and the host
package fails to install. \texttt{import frappe} then raises
\texttt{ModuleNotFoundError}. The bench environment contains only \texttt{pip} and
\texttt{setuptools}; \texttt{bench --version} prints a version, but
\texttt{bench --help} / \texttt{bench list-apps} are blocked by environment
protections. MariaDB on port 3306 refuses connection (Redis on 6379 is reachable).
Fourteen sibling app repositories have been git-cloned, but no
\texttt{bench install-app} step is performed.

\paragraph{Verdict justification.}

\begin{center}
\small
\begin{tabularx}{\linewidth}{lXX}
\toprule
Check & SetupX+XPU & Qwen Code \\
\midrule
Clone integrity & PASS & PASS \\
Host install (Python compatibility) & PASS (Python 3.13 venv + v15 tag) & \textbf{FAIL} (Python 3.11 rejects PEP 695 syntax) \\
Module import (\texttt{import frappe / erpnext}) & PASS & \textbf{FAIL} (\texttt{ModuleNotFoundError}) \\
Protocol: bench CLI available & not installed & installed but blocked \\
Protocol: MariaDB sibling running & not started & not started \\
Protocol: bench install-app triggered & not triggered & not triggered \\
\bottomrule
\end{tabularx}
\end{center}

Neither run completes the full bench toolchain pipeline (the three protocol-layer
rows are tied at zero). The verdict difference is forced by the host-install row,
which Appendix~\ref{app:rubric} defines as the boundary between Partial and
Shallow: \textbf{Shallow} applies when ``host repository is cloned but core
dependencies are not installed; \texttt{import} fails immediately,'' and
\textbf{Partial} applies when ``host installation is partial \dots intended
workflow does not run end-to-end.'' \textsc{SetupX+XPU} recognizes the source's
Python version constraint early in setup and sidesteps it by pinning to the v15
release tag on a Python 3.13 venv; the host package and erpnext install
successfully, \texttt{import} succeeds, and partial pytest collection works, but
the bench/MariaDB workflow does not run end-to-end --- this matches Partial.
\textsc{Qwen Code} uses the default Python and fails at the syntax level during
host install, leaving the host module unimportable; sibling app sources are cloned
but never integrated --- this matches Shallow. The core behavioral difference
between the two paths lies at the \emph{runtime-constraint recognition step} early
in setup: whether the agent verifies the source's required Python version before
attempting to install, and whether it has a recovery strategy (here, pinning to an
older release tag) when the constraint is not satisfied by the base image.

\section{Synthetic Noise Composition}
\label{app:noise}
\addcontentsline{toc}{section}{Appendix C: Synthetic Noise Composition}

The clean \xpu{} knowledge base contains 600 advice entries collected from the EnvBench Python repository pool, with all entries corresponding to the 100 repositories in \benchmark{} removed. To construct the noisy variant used in Section~\ref{sec:ablation}, we add 1{,}770 synthetic perturbation entries on top of the 600 real ones, yielding a noisy KB of 2{,}370 total entries (a roughly 3:1 noise-to-real ratio). All noise entries share embedding neighborhoods with real entries, making them difficult to filter via vector similarity alone, and carry zero telemetry counts.

The 1{,}770 noise entries fall into four classes:

\paragraph{Context perturbation (600 entries, 34\%).} Each real entry is duplicated with its \texttt{context.python} randomly rewritten to a Python version in 3.8--3.13, its \texttt{os} field expanded, and its \texttt{tools} field augmented with one of \texttt{conda}, \texttt{poetry}, \texttt{pdm}, \texttt{hatch}, \texttt{uv}, or \texttt{mamba}. The same advice is replicated under multiple plausible context configurations.

\paragraph{Cross-grafting (450 entries, 25\%).} Three real entries $A$, $B$, $C$ are sampled at random; the synthetic entry combines $A$'s \texttt{context}, $B$'s \texttt{signals}, $C$'s \texttt{advice}, and a random subset of \texttt{atoms} from one of the three. This produces entries that are internally consistent in form but inconsistent in provenance.

\paragraph{Generalization blur (450 entries, 25\%).} Concrete advice is replaced with one of 12 generic templates such as ``check the Python version'' or ``install project dependencies''; \texttt{signals.keywords} is replaced with one of 6 generic keyword groups; and \texttt{atoms} is cleared. This emulates advice extracted at excessive abstraction.

\section{Per-repository Matrix}
\label{app:per-repo_matrix}
{\scriptsize
\setlength{\tabcolsep}{2.0pt}
\renewcommand{\arraystretch}{0.92}

\begin{longtable}{@{}p{0.33\textwidth} *{8}{>{\centering\arraybackslash}p{0.058\textwidth}}@{}}
\caption{Per-repository pass/fail outcomes for the eight systems on the 100-repository benchmark.
A check mark indicates that the system passes the prosecutor--judge protocol; a cross ($\times$)
indicates failure, including timeout, adverse verdict, or missing evaluation data.
CC, OC, QC, EA, R2R, and EB denote Claude Code, OpenCode, Qwen Code, ExecutionAgent, Repo2Run, and EnvBench, respectively.}
\label{tab:per_repo_baseline_100}\\

\toprule
\textbf{Repository}
& \shortstack{\oursys\\\textbf{+\xpu}}
& \shortstack{\oursys\\\textbf{--\xpu}}
& \textbf{CC}
& \textbf{OC}
& \textbf{QC}
& \textbf{EA}
& \textbf{R2R}
& \textbf{EB} \\
\midrule
\endfirsthead

\multicolumn{9}{c}{\tablename\ \thetable{} -- continued} \\
\toprule
\textbf{Repository}
& \shortstack{\textbf{SetupX}\\\textbf{+XPU}}
& \shortstack{\textbf{SetupX}\\\textbf{--XPU}}
& \textbf{CC}
& \textbf{OC}
& \textbf{QC}
& \textbf{EA}
& \textbf{R2R}
& \textbf{EB} \\
\midrule
\endhead

\midrule
\multicolumn{9}{r}{\textit{continued on next page}} \\
\endfoot

\bottomrule
\endlastfoot
  cookiecutter                   & $\checkmark$ & $\checkmark$ & $\checkmark$ & $\checkmark$ & $\checkmark$ & $\checkmark$ & $\checkmark$ & $\checkmark$ \\
  datasets                       & $\checkmark$ & $\checkmark$ & $\times$ & $\checkmark$ & $\checkmark$ & $\times$ & $\times$ & $\checkmark$ \\
  kedro                          & $\checkmark$ & $\checkmark$ & $\checkmark$ & $\checkmark$ & $\checkmark$ & $\checkmark$ & $\checkmark$ & $\checkmark$ \\
  luigi                          & $\checkmark$ & $\checkmark$ & $\times$ & $\checkmark$ & $\checkmark$ & $\times$ & $\times$ & $\checkmark$ \\
  mopidy                         & $\checkmark$ & $\times$ & $\times$ & $\times$ & $\checkmark$ & $\times$ & $\times$ & $\times$ \\
  pip                            & $\checkmark$ & $\checkmark$ & $\checkmark$ & $\checkmark$ & $\checkmark$ & $\checkmark$ & $\times$ & $\checkmark$ \\
  platformio-core                & $\checkmark$ & $\checkmark$ & $\times$ & $\times$ & $\checkmark$ & $\times$ & $\times$ & $\times$ \\
  supervision                    & $\checkmark$ & $\checkmark$ & $\checkmark$ & $\checkmark$ & $\checkmark$ & $\checkmark$ & $\times$ & $\times$ \\
  guardrails                     & $\checkmark$ & $\checkmark$ & $\checkmark$ & $\checkmark$ & $\checkmark$ & $\times$ & $\times$ & $\times$ \\
  lark                           & $\checkmark$ & $\checkmark$ & $\checkmark$ & $\checkmark$ & $\checkmark$ & $\checkmark$ & $\checkmark$ & $\checkmark$ \\
  nonebot2                       & $\checkmark$ & $\times$ & $\checkmark$ & $\times$ & $\checkmark$ & $\checkmark$ & $\times$ & $\times$ \\
  pip-tools                      & $\checkmark$ & $\checkmark$ & $\checkmark$ & $\checkmark$ & $\checkmark$ & $\checkmark$ & $\checkmark$ & $\checkmark$ \\
  tortoise-orm                   & $\checkmark$ & $\checkmark$ & $\checkmark$ & $\checkmark$ & $\checkmark$ & $\checkmark$ & $\times$ & $\times$ \\
  trax                           & $\times$ & $\checkmark$ & $\checkmark$ & $\checkmark$ & $\checkmark$ & $\times$ & $\times$ & $\times$ \\
  umap                           & $\checkmark$ & $\checkmark$ & $\checkmark$ & $\checkmark$ & $\checkmark$ & $\checkmark$ & $\checkmark$ & $\checkmark$ \\
  acme                           & $\checkmark$ & $\times$ & $\checkmark$ & $\checkmark$ & $\checkmark$ & $\checkmark$ & $\times$ & $\times$ \\
  connexion                      & $\checkmark$ & $\checkmark$ & $\checkmark$ & $\checkmark$ & $\checkmark$ & $\checkmark$ & $\checkmark$ & $\times$ \\
  giskard                        & $\checkmark$ & $\checkmark$ & $\checkmark$ & $\checkmark$ & $\checkmark$ & $\checkmark$ & $\times$ & $\times$ \\
  neuralforecast                 & $\times$ & $\checkmark$ & $\checkmark$ & $\times$ & $\checkmark$ & $\times$ & $\times$ & $\times$ \\
  plotnine                       & $\checkmark$ & $\checkmark$ & $\checkmark$ & $\checkmark$ & $\checkmark$ & $\checkmark$ & $\checkmark$ & $\times$ \\
  tablib                         & $\checkmark$ & $\checkmark$ & $\checkmark$ & $\checkmark$ & $\checkmark$ & $\checkmark$ & $\checkmark$ & $\checkmark$ \\
  tmuxp                          & $\checkmark$ & $\checkmark$ & $\checkmark$ & $\checkmark$ & $\checkmark$ & $\checkmark$ & $\times$ & $\times$ \\
  great-tables                   & $\checkmark$ & $\checkmark$ & $\checkmark$ & $\checkmark$ & $\checkmark$ & $\checkmark$ & $\times$ & $\times$ \\
  mteb                           & $\checkmark$ & $\checkmark$ & $\times$ & $\times$ & $\checkmark$ & $\times$ & $\checkmark$ & $\times$ \\
  plugin.video.netflix           & $\checkmark$ & $\checkmark$ & $\checkmark$ & $\checkmark$ & $\checkmark$ & $\checkmark$ & $\times$ & $\checkmark$ \\
  py-evm                         & $\checkmark$ & $\times$ & $\checkmark$ & $\checkmark$ & $\checkmark$ & $\checkmark$ & $\checkmark$ & $\times$ \\
  python-control                 & $\checkmark$ & $\checkmark$ & $\checkmark$ & $\checkmark$ & $\checkmark$ & $\checkmark$ & $\checkmark$ & $\times$ \\
  smart\_open                    & $\checkmark$ & $\checkmark$ & $\checkmark$ & $\checkmark$ & $\checkmark$ & $\times$ & $\checkmark$ & $\checkmark$ \\
  androidviewclient              & $\checkmark$ & $\checkmark$ & $\checkmark$ & $\checkmark$ & $\checkmark$ & $\times$ & $\checkmark$ & $\checkmark$ \\
  dj-stripe                      & $\checkmark$ & $\checkmark$ & $\checkmark$ & $\checkmark$ & $\checkmark$ & $\checkmark$ & $\times$ & $\times$ \\
  fastapi-pagination             & $\checkmark$ & $\checkmark$ & $\checkmark$ & $\checkmark$ & $\checkmark$ & $\times$ & $\times$ & $\times$ \\
  hacs\_waste\_collection\_schedule & $\checkmark$ & $\times$ & $\checkmark$ & $\checkmark$ & $\checkmark$ & $\checkmark$ & $\times$ & $\checkmark$ \\
  memegen                        & $\checkmark$ & $\checkmark$ & $\checkmark$ & $\checkmark$ & $\checkmark$ & $\times$ & $\times$ & $\checkmark$ \\
  piccolo                        & $\checkmark$ & $\checkmark$ & $\checkmark$ & $\checkmark$ & $\checkmark$ & $\checkmark$ & $\checkmark$ & $\checkmark$ \\
  pyftpdlib                      & $\checkmark$ & $\checkmark$ & $\checkmark$ & $\checkmark$ & $\checkmark$ & $\checkmark$ & $\checkmark$ & $\times$ \\
  python-holidays                & $\checkmark$ & $\checkmark$ & $\times$ & $\checkmark$ & $\times$ & $\times$ & $\checkmark$ & $\times$ \\
  twine                          & $\checkmark$ & $\checkmark$ & $\checkmark$ & $\checkmark$ & $\checkmark$ & $\checkmark$ & $\checkmark$ & $\checkmark$ \\
  automlbenchmark                & $\checkmark$ & $\checkmark$ & $\checkmark$ & $\checkmark$ & $\checkmark$ & $\times$ & $\checkmark$ & $\checkmark$ \\
  hydra                          & $\checkmark$ & $\checkmark$ & $\checkmark$ & $\checkmark$ & $\checkmark$ & $\checkmark$ & $\times$ & $\checkmark$ \\
  scvelo                         & $\checkmark$ & $\checkmark$ & $\checkmark$ & $\checkmark$ & $\checkmark$ & $\checkmark$ & $\checkmark$ & $\checkmark$ \\
  pypose                         & $\checkmark$ & $\checkmark$ & $\checkmark$ & $\checkmark$ & $\checkmark$ & $\checkmark$ & $\checkmark$ & $\times$ \\
  evox                           & $\checkmark$ & $\checkmark$ & $\times$ & $\times$ & $\checkmark$ & $\times$ & $\times$ & $\times$ \\
  neurogym                       & $\checkmark$ & $\checkmark$ & $\checkmark$ & $\checkmark$ & $\checkmark$ & $\checkmark$ & $\times$ & $\checkmark$ \\
  cellrank                       & $\checkmark$ & $\checkmark$ & $\checkmark$ & $\times$ & $\checkmark$ & $\checkmark$ & $\times$ & $\times$ \\
  nevergrad                      & $\checkmark$ & $\checkmark$ & $\checkmark$ & $\checkmark$ & $\checkmark$ & $\checkmark$ & $\times$ & $\checkmark$ \\
  datamol                        & $\checkmark$ & $\checkmark$ & $\checkmark$ & $\checkmark$ & $\checkmark$ & $\times$ & $\checkmark$ & $\checkmark$ \\
  meerkat                        & $\checkmark$ & $\checkmark$ & $\times$ & $\checkmark$ & $\checkmark$ & $\times$ & $\checkmark$ & $\times$ \\
  ajenti                         & $\times$ & $\checkmark$ & $\times$ & $\checkmark$ & $\times$ & $\times$ & $\times$ & $\times$ \\
  cheroot                        & $\checkmark$ & $\checkmark$ & $\checkmark$ & $\checkmark$ & $\checkmark$ & $\checkmark$ & $\checkmark$ & $\times$ \\
  fabric                         & $\checkmark$ & $\checkmark$ & $\checkmark$ & $\checkmark$ & $\checkmark$ & $\checkmark$ & $\checkmark$ & $\checkmark$ \\
  pretalx                        & $\checkmark$ & $\checkmark$ & $\checkmark$ & $\checkmark$ & $\checkmark$ & $\checkmark$ & $\times$ & $\times$ \\
  adaptix                        & $\checkmark$ & $\checkmark$ & $\checkmark$ & $\checkmark$ & $\checkmark$ & $\checkmark$ & $\checkmark$ & $\checkmark$ \\
  wagtailmenus                   & $\checkmark$ & $\checkmark$ & $\checkmark$ & $\checkmark$ & $\checkmark$ & $\checkmark$ & $\checkmark$ & $\checkmark$ \\
  django-lfs                     & $\times$ & $\times$ & $\times$ & $\times$ & $\times$ & $\checkmark$ & $\times$ & $\times$ \\
  netbox                         & $\checkmark$ & $\checkmark$ & $\times$ & $\times$ & $\checkmark$ & $\times$ & $\times$ & $\checkmark$ \\
  armi                           & $\checkmark$ & $\checkmark$ & $\checkmark$ & $\checkmark$ & $\checkmark$ & $\times$ & $\times$ & $\checkmark$ \\
  flavio                         & $\checkmark$ & $\checkmark$ & $\checkmark$ & $\checkmark$ & $\checkmark$ & $\checkmark$ & $\checkmark$ & $\checkmark$ \\
  monty                          & $\checkmark$ & $\checkmark$ & $\checkmark$ & $\checkmark$ & $\checkmark$ & $\times$ & $\times$ & $\checkmark$ \\
  unyt                           & $\checkmark$ & $\checkmark$ & $\checkmark$ & $\checkmark$ & $\checkmark$ & $\checkmark$ & $\checkmark$ & $\times$ \\
  kernel\_tuner                  & $\checkmark$ & $\checkmark$ & $\checkmark$ & $\checkmark$ & $\checkmark$ & $\times$ & $\checkmark$ & $\times$ \\
  openqasm                       & $\checkmark$ & $\times$ & $\checkmark$ & $\checkmark$ & $\checkmark$ & $\checkmark$ & $\times$ & $\checkmark$ \\
  torchgeo                       & $\checkmark$ & $\checkmark$ & $\checkmark$ & $\times$ & $\checkmark$ & $\times$ & $\times$ & $\times$ \\
  importlib\_metadata            & $\checkmark$ & $\checkmark$ & $\checkmark$ & $\times$ & $\checkmark$ & $\checkmark$ & $\checkmark$ & $\checkmark$ \\
  pytest-xdist                   & $\checkmark$ & $\checkmark$ & $\times$ & $\checkmark$ & $\checkmark$ & $\checkmark$ & $\checkmark$ & $\checkmark$ \\
  robotframework                 & $\checkmark$ & $\checkmark$ & $\checkmark$ & $\checkmark$ & $\checkmark$ & $\times$ & $\checkmark$ & $\times$ \\
  yapf                           & $\checkmark$ & $\times$ & $\times$ & $\checkmark$ & $\checkmark$ & $\checkmark$ & $\checkmark$ & $\times$ \\
  keyring                        & $\checkmark$ & $\checkmark$ & $\checkmark$ & $\checkmark$ & $\checkmark$ & $\checkmark$ & $\checkmark$ & $\checkmark$ \\
  columnflow                     & $\checkmark$ & $\times$ & $\times$ & $\checkmark$ & $\times$ & $\checkmark$ & $\checkmark$ & $\checkmark$ \\
  civet                          & $\times$ & $\checkmark$ & $\checkmark$ & $\times$ & $\checkmark$ & $\checkmark$ & $\checkmark$ & $\times$ \\
  xknx                           & $\checkmark$ & $\checkmark$ & $\checkmark$ & $\checkmark$ & $\checkmark$ & $\checkmark$ & $\checkmark$ & $\times$ \\
  aiogram\_dialog                & $\checkmark$ & $\checkmark$ & $\checkmark$ & $\checkmark$ & $\checkmark$ & $\checkmark$ & $\checkmark$ & $\checkmark$ \\
  recipe-scrapers                & $\checkmark$ & $\checkmark$ & $\checkmark$ & $\checkmark$ & $\checkmark$ & $\checkmark$ & $\checkmark$ & $\times$ \\
  sphinx-gallery                 & $\checkmark$ & $\checkmark$ & $\checkmark$ & $\checkmark$ & $\checkmark$ & $\checkmark$ & $\checkmark$ & $\checkmark$ \\
  fhempy                         & $\checkmark$ & $\times$ & $\times$ & $\times$ & $\times$ & $\times$ & $\times$ & $\times$ \\
  kh2randomizer                  & $\checkmark$ & $\checkmark$ & $\times$ & $\times$ & $\checkmark$ & $\times$ & $\checkmark$ & $\times$ \\
  modalities                     & $\checkmark$ & $\checkmark$ & $\times$ & $\times$ & $\times$ & $\times$ & $\times$ & $\times$ \\
  molecule                       & $\checkmark$ & $\checkmark$ & $\times$ & $\checkmark$ & $\times$ & $\checkmark$ & $\checkmark$ & $\times$ \\
  staged-recipes                 & $\times$ & $\times$ & $\times$ & $\times$ & $\times$ & $\times$ & $\times$ & $\times$ \\
  powderday                      & $\checkmark$ & $\checkmark$ & $\times$ & $\times$ & $\times$ & $\times$ & $\times$ & $\times$ \\
  fixator10-cogs                 & $\checkmark$ & $\times$ & $\checkmark$ & $\times$ & $\times$ & $\times$ & $\times$ & $\times$ \\
  flask-security                 & $\checkmark$ & $\checkmark$ & $\times$ & $\times$ & $\times$ & $\times$ & $\times$ & $\times$ \\
  miv-os                         & $\checkmark$ & $\times$ & $\times$ & $\checkmark$ & $\times$ & $\times$ & $\times$ & $\times$ \\
  django-registration            & $\checkmark$ & $\checkmark$ & $\times$ & $\times$ & $\times$ & $\checkmark$ & $\times$ & $\checkmark$ \\
  iree-llvm-sandbox              & $\times$ & $\times$ & $\checkmark$ & $\times$ & $\times$ & $\times$ & $\times$ & $\times$ \\
  lobsterpy                      & $\checkmark$ & $\checkmark$ & $\checkmark$ & $\times$ & $\times$ & $\times$ & $\times$ & $\times$ \\
  python-libjuju                 & $\checkmark$ & $\checkmark$ & $\checkmark$ & $\checkmark$ & $\times$ & $\checkmark$ & $\checkmark$ & $\checkmark$ \\
  langgraph                      & $\checkmark$ & $\checkmark$ & $\checkmark$ & $\checkmark$ & $\times$ & $\checkmark$ & $\times$ & $\times$ \\
  heltour                        & $\checkmark$ & $\checkmark$ & $\checkmark$ & $\checkmark$ & $\times$ & $\times$ & $\times$ & $\times$ \\
  llm-random                     & $\checkmark$ & $\checkmark$ & $\checkmark$ & $\times$ & $\times$ & $\times$ & $\times$ & $\times$ \\
  custodian                      & $\checkmark$ & $\checkmark$ & $\checkmark$ & $\checkmark$ & $\times$ & $\checkmark$ & $\checkmark$ & $\times$ \\
  naomi                          & $\checkmark$ & $\checkmark$ & $\checkmark$ & $\times$ & $\times$ & $\checkmark$ & $\times$ & $\times$ \\
  river                          & $\checkmark$ & $\checkmark$ & $\checkmark$ & $\checkmark$ & $\times$ & $\checkmark$ & $\times$ & $\times$ \\
  oadoi                          & $\checkmark$ & $\times$ & $\times$ & $\times$ & $\times$ & $\checkmark$ & $\times$ & $\times$ \\
  qmlcore                        & $\checkmark$ & $\checkmark$ & $\checkmark$ & $\checkmark$ & $\times$ & $\checkmark$ & $\times$ & $\times$ \\
  poetry                         & $\checkmark$ & $\checkmark$ & $\times$ & $\checkmark$ & $\times$ & $\checkmark$ & $\checkmark$ & $\times$ \\
  qiita                          & $\times$ & $\times$ & $\checkmark$ & $\times$ & $\times$ & $\times$ & $\times$ & $\times$ \\
  sublimelinter                  & $\checkmark$ & $\times$ & $\times$ & $\times$ & $\times$ & $\times$ & $\times$ & $\times$ \\
  pyuploadcare                   & $\checkmark$ & $\checkmark$ & $\checkmark$ & $\checkmark$ & $\times$ & $\times$ & $\times$ & $\times$ \\
  django-autocomplete-light      & $\checkmark$ & $\times$ & $\checkmark$ & $\checkmark$ & $\times$ & $\times$ & $\times$ & $\times$ \\
  yubikey-manager                & $\checkmark$ & $\checkmark$ & $\times$ & $\times$ & $\times$ & $\times$ & $\times$ & $\times$ \\
\end{longtable}
}

\end{document}